# Heterogeneous Responses to the U.S. Narrative Tax Changes: Evidence from the U.S. States

Masud Alam[1]

## Abstract:

This paper investigates the assumption of homogeneous effects of federal tax changes across the U.S. states and identifies where and why that assumption may not be valid. More specifically, what determines the transmission mechanism of tax shocks at the state level? How vital are states' fiscal structures, financial conditions, labor market rigidities, and industry mix? Do these economic and structural characteristics drive the transmission mechanism of the tax changes at the state level at different horizons? This study employs a panel factor-augmented vector autoregression (FAVAR) technique to answer these issues. The findings show that state economies respond homogeneously in terms of employment and price levels; however, they react heterogeneously in real GDP and personal income growth. In most states, these reactions are statistically significant, and the heterogeneity in the effects of tax cuts is significantly related to the state's fiscal structure, manufacturing and financial composition, and the labor market's rigidity. A cross-state regression analysis shows that states with higher tax elasticity, higher personal income tax, strict labor market regulation, and economic policy uncertainties are relatively less responsive to federal tax changes. In contrast, the magnitude of the response in real GDP, personal income, and employment to tax cuts is relatively higher in states with a larger share of finance, manufacturing, lower tax burdens, and flexible credit markets.

***Key words:*** Fiscal policy, Narrative tax changes, U.S. states, Heterogeneous responses, FAVAR model, tax shocks, Real GDP.

***JEL Codes:*** E2, E3, E6, E17, H3, H7.

[1] Northern Illinois University, IL, USA, & Shahjalal University of Science and Technology, Bangladesh. Email: masud.sust@gmail.com

# Heterogeneous Responses to the U.S. Narrative Tax Changes: Evidence from the U.S. States

Masud Alam[1]

## 1. Introduction

The existing literature examining the macroeconomic effects of federal tax changes assume that tax shocks affect all states the same way, which does not seem realistic. The empirical literature on the impact of federal tax changes, pioneered by Romer and Romer (2010), and later Alam (2021a) and Mertens and Ravn (2013), uses narrative tax shocks to estimate an aggregate tax multiplier. However, this aggregate estimate does not consider the importance of state-level fiscal capacities, industrial mix, financial frictions, and labor market heterogeneities in the transmission of federal tax shocks.

While all states are subject to identical federal tax policy, every state has a unique mix of policy and political ideology impacting output growth, employment, and investment. States are also characterized by different public institutions, economic structures, and levels of regulations controlling public debt. Such diversity and policy complexity across the states' highlights the importance of examining the validity of the homogeneous impact assumption and, if found unwarranted, to uncover what factors within the state economies impacted and how.

A number of recent literature (Carlino and Defina, 1998; Hussain and Malik, 2016; Liu and Williams, 2019; Owyang and Zubairy, 2013) has incorporated this state heterogeneities in small scale SVAR models. They examine the impacts of federal policy changes on state-level economic activities. Estimation of a small scale SVAR model in these state-level studies, however, may omit information sets associated with regional and federal economic variables. This omitted variable bias frequently led to forecasts that run contrary to the standard macroeconomic theory, such as price puzzles and insignificant impulse responses.

Additionally, the measurement issue of omitted variable bias can be a serious problem when we do not have theoretical criteria about the number and the choice of relevant states and regional variables. As Mertens (2019) mentions, the highly endogenous nature of federal tax changes implies that the small-scale VAR model with only state-level macroeconomic variables cannot capture sufficient information. The information insufficiency could be the potential reason to expect that the estimates are biased, and the confidence intervals are relatively wider than assumed. The dynamics of macroeconomic variables across the U.S. states are possibly related to a far wider regional and aggregate policy variables than typically considered in small-scale VAR models. Therefore, the estimation of federal tax policy changes is likely to be biased if the additional regional and aggregate information is not included in the VAR system. Several alternative specification schemes ranging from Bayesian VAR models to sign restrictions and the penalty function approach are proposed and applied to solve the identification problems (Banbura, Giannone and Reichlin, 2007; Scholl and Uhlig, 2006; Uhlig, 2005). Depending on the choice of a suitable prior and the selection of appropriate a priori signs, the forecast analysis with these methods may produce mixed results.

[1] Northern Illinois University, USA & Shahjalal University of Science and Technology, Bangladesh. Email: masud.sust@gmail.com.

Given the disaggregated level focus of this study, a state-specific small-scale structural VAR model may deliver biased estimates (Hansen and Sargent, 1991; Mertens, 2019; Stock and Watson, 2018), which could be solved with additional control variables. This study attempts to solve the identification and dimensionality problems with structural VAR models by estimating a state-level factor augmented vector autoregressive (FAVAR) model across all 50 states. The model extracts common U.S. regional and aggregate macroeconomic factors from an extensive panel of 48 economic variables covering eight Bureau of Economic Analysis (BEA) regions and a panel of 132 federal-level output, prices, employment, and monetary and financial variables.

The FAVAR model draws information from the BEA regional variables and the comovement of the aggregate of macroeconomic variables and confirms that the dynamics of federal and regional variables are adequately controlled through the factors. The specific technique allows this study to deal with dimensionality reduction of the federal- or region-specific unobserved influences while mitigating the omitted variable bias in a parsimonious way. As a result, the magnitude and the direction of impulse response functions that the specification delivers through the control of factors ensure that the state-level estimates are likely unbiased and significantly similar to the federal counterparts.

The fundamental concern in estimating the effect of federal tax changes on state-level economic activity is that tax changes are endogenous, where variation in tax policies are directly correlated with contemporary economic conditions. To mitigate the endogeneity problem, this study adopts the narrative approach (Alam, 2021a; Romer and Romer, 2009) to identify exogenous tax changes by relying on the reports published by Congressional Budget Office, the Committee on Ways and Means, Senate and Congressional records, and the Economic Report of the President. This method identifies tax changes that are designed to stimulate long-run growth and are not directly related to current economic conditions. The narrative dataset of this study includes eleven personal income and nine corporate income tax changes, which are used in the FAVAR models by imposing sign restrictions on the responses of state-level macroeconomic variables with Uhlig's (2005) penalty function to estimate the impulse response functions (IRFs). The penalty function ensures that policy shocks are orthogonal to other shocks and deliver a unique solution for the impulse response functions.

The FAVAR estimates suggest that the magnitude and persistence of shocks to GDP, employment, personal income, and price responses are heterogeneous across states, and, in most states, these responses are statistically and economically significant. The impulse response functions (IRFs) support the imposed a priori signs from the theoretical prediction of the New Keynesian model of an unanticipated tax cut. A one percent cut in personal income tax increases real GDP by about 1.2 percent on impact and a maximum of 1.9 percent after four years. A one percent cut in corporate income tax cut raises real GDP immediately by 0.52 percent and by a maximum of 0.83 percent after two years. Cuts in personal income tax increase real GDP in 33 states, raise personal income in 39 states and drives the price level up in 46 states. The same unanticipated corporate tax cuts produce a statistically significant rise in real GDP and personal income for around 40 states and price level rises in 48 states. For all other states, the effect of tax cuts are either contractionary or insignificant. The direction of non-farm employment response is significantly homogenous in regard to the shock across the U.S. states; however, the magnitude and persistence of employment responses are most considerable in 14 states, with employment rising by a maximum of about 1.6 percent after the personal income tax cut and 0.8 percent for the corporate tax cut. The impact of either tax cut persists for more than five years.

The study then investigates the importance of state-specific structural characteristics for the transmission mechanism and the heterogeneous responses of federal tax changes. A cross-state regression analysis between state-specific cumulative responses and state structural characteristics shows that states with a higher financial and manufacturing industry share, smaller tax burdens, smaller degree of labor market rigidities, less economic policy uncertainty, and a higher capacity of job creation appear to be the more responsive states in real GDP and personal income to the change in either tax. In contrast, the magnitude of the response in price levels and employment to a corporate tax cut is relatively higher in states with broader credit channels, a smaller degree of financial friction, and a smaller degree of labor market rigidities. Results further show that the estimated average response is higher for a personal income tax cut than the same of cut in corporate income tax.  These empirical findings are broadly consistent with recent disaggregated studies by Auerbach (2006); Gravelle (2011), Owyang and Zubairy (2013), and Suarez and Zidar (2016).

## 2. Literature review

Literature examining the effects of tax changes at the state level is relatively vast and generally divided into two competing categories. The first category follows the microeconomic approach (as in Giroud and Rauh, 2019; Ljungqvist and Smolyansky, 2014; Serrato and Zidar, 2018), mainly focuses on cross-states comparisons. While the others follow the second, or macroeconomic approach, mostly relying on small-scale SVAR models (Hooker and Knetter, 1997; Hussain and Malik, 2016; Liu and Williams, 2019; Nekarda and Ramey, 2011; Owyang and Zubairy, 2013), with a recent application of the FAVAR approach in Herrera and Rangaraju (2019).

Most of the micro-based literature focusing on the state-level effects of tax policy changes can be categorized into two: those that examine the impact on output growth and those that examine employment effects of tax rate changes. The literature examining the impact of state-level tax changes on output growth beginning with Helms (1985) and Mullen and William (1994), per capita income growth in Holcombe and Lacombe (2004) and Reed and Rogers (2004), and the robust relationship between tax cut and income growth in the U.S. states by Reed (2008) and Poulson and Kaplan (2008).

Helms (1985) and Mullen and William (1994) both study the effects of state and local tax increase on economic growth. They find that higher marginal tax rates significantly impede state-level output growth. Holcombe and Lacombe (2004) compare the effects of income tax rates on the state border county's per capita income growth with the neighboring state counties. They conclude that states with higher income taxes experience negative income growth. On average, findings show a 3.4 percent decrease in per capita income growth for states that increases their marginal personal income tax compared to their neighboring states.

Goff, Lebedinsky, and Lile (2012) examine the effects of state-level tax rates on state-level per-capita output growth by employing a pairwise matching technique. They argue that the pairwise states-matching approach is better to control state-level omitted variables bias. Their findings suggest a growth-enhancing effect of lower tax rates, where a one percentage point increase in tax burden reduces state-level output growth by a maximum of 2.19 percentage point. A variant approach is used by Bania, Gray, and Stone (2007), who study the non-monotonic effects of U.S. state taxes on economic growth. They employ a Barro-style endogenous growth model that also shows that the economic growth declines with an increase in state-level tax rates.

In addition to overall economic growth, Gale, Krupkin and Rueben, (2015) extend the model in Reed (2008) and investigate how state tax policy affects employment and entrepreneurial activity across states and over time. Contrary to Reed's findings, they find cuts in top state marginal tax rates do not necessarily generate growth rates across states. In a recent study, Zidar (2019) finds the distribution of income across U.S. states strongly correlated with the variation in the response of federal tax cuts, which is similar to the cross-states regression results of this study, where both median income and the state's tax burden have significant effects on state-level responses. Although there are some exceptions (as in Ahamed, 2021a; Gale, Krupkin and Rueben, 2015; Gruber and Saez, 2002; Yang, 2005), a substantial consensus in the literature suggests the robust relationships between a tax cut and economic growth across U.S. states.

On the employment effects of tax policy changes, Mark, McGuire, and Papke (2000) estimate the employment effects of personal, sales, and business property taxes on employment growth in Washington, DC. They find that increase in the personal property tax rate by one percentage point reduces employment growth by a maximum of 2.44 percentage points; however, there are no significant effects of corporate income tax rates on employment growth. For New Jersey county-level data, Reed and Rogers (2004) find a strong employment growth response to a 30 percent cut in personal income tax than the counties in neighboring states that maintain same tax rates or increase taxes.

In a unique contribution to the corporate tax literature, Ljungqvist and Smolyansky (2014) examine the impact of corporate taxes on employment and income across U.S. states and investigate the role that the economic recession may play in the employment effects. They employ a border approach on resident-based employment and income data by comparing counties located across state borders. The assumption is that the corporate tax policy changes in one state and not the other. Findings show that an increase in the top marginal corporate tax rate by one-percentage-point reduces employment by 0.3–0.5 percent and income by 0.3-0.5 percentage points. Following similar methodology, Shuai and Chmura (2013) compare states that reduce corporate income tax rates with those making no changes. They also find the positive impact of corporate tax cuts on faster job creation.

While only a small number of literatures reaches conflicting conclusions on the employment effects of corporate tax cuts, other related studies suggest that reducing personal and corporate tax rates results in increased state-level employment and investment. In the most recent paper, Garrett, Ohrn, and Serrato (2020) examine the stimulus effects of federal corporate tax policy on county-level employment and earnings. They find counties that experience a larger decrease in corporate tax-induced investment costs see an increase in employment by a maximum of 1.9 percent. Goss and Philips (1994) find evidence that higher personal income taxes reduce state-level employment growth, but not for the reduction of corporate income taxes.

In macroeconomic approach where SVAR models predominantly employed, the main advantage is that it exploits information directly from the data. The SVAR technique, however, requires researchers to impose *a priori* restrictions on the model parameters and can employ only a limited information set to model in examining the effects of the policy changes, typically yielding counter-intuitive empirical results (Ahamed, 2021b; Forni et al., 2005; Hansen and Sargent, 1991; Stock and Watson, 2016). Another concern with SVAR is the identification of a structural shock that may not purely orthogonal to other shocks in the model. Various restrictions and approaches have been proposed in the empirical literature to overcome this concern. For example, conventional restrictions include short- and long-run restrictions on the structural shocks (Hua, Polansky, and Pramanik, 2019; Blanchard and Quah, 1989; Kim and Roubini, 2000; Polansky and Pramanik, 2021; Pramanik, 2020; Pramanik and Polansky, 2021; Rapach, 2001), sign restrictions (Faust, 1998; Fry and Pagan, 2011; and Uhlig, 2005), external instrument approach (Hossain and Ahamed, 2015; Romer and Romer, 2004, 2009; Mertens and Ravn, 2012; and Stock and Watson, 2012), or combine any two approaches as in Braun and Brüggemann (2017) and Ludvigson, Ma, and Ng (2017).

To overcome the difficulties in identifying exogenous tax shocks, some recent studies rely on exogenous instrumental measures on the state-level heterogeneous responses, such as Owyang and Zubairy (2013), who examine the aggregate effects of military spending shocks on state-level personal income and employment and uncover the asymmetry in regional patterns. Nekarda and Ramey (2011) show heterogeneous effects of government spending on output and labor market variables across industries. Hooker and Knetter (1997) also find greater variation of employment growth across states to military procurement spending shock. Hussain and Malik (2016) examine the asymmetric effects of unanticipated tax changes on states' output using a non-linear VAR approach. They find asymmetric non-linear impulse responses only for personal income tax changes, whereas a considerable symmetry appears in output response to the corporate income tax changes. These studies are quite similar in the spirit of exogenous instrumental SVAR approach that use non-tax changes to proxy for tax changes, such as government spending, military spending, and total revenue changes, and mostly estimating low-dimensional VAR models.

An alternative to SVAR models, FAVAR approach deals the dimensionality issues parsimoniously and most suited to examine policy effectiveness when underlying policy variables are unobserved. The FAVAR estimates a few unobserved latent factors from a large panel of economic dataset that better represent the entire economy's aggregate comovement. This small number of latent factors minimizes the risk of priori variable selection or *ad-hoc* decisions about the number of variables related to SVAR models. While the FAVAR approach has primarily been used to identify monetary policy shocks, the empirical application on U.S. fiscal policy shocks is quite limited. Examples of such studies include Bénassy-Quéré and Cimadomo (2006), Forni and Gambetti (2010), and Herrera and Rangaraju (2019).

Bénassy-Quéré and Cimadomo (2006) and Forni and Gambetti (2010) apply dynamic FAVAR models to estimate fiscal policy multipliers in the U.S. and the European countries. In the former, the authors use government spending and net tax shocks impacting the cross-border fiscal spillovers from Germany to other largest European economies while controlling for three global factors in a VAR model. In the latter, Forni and Gambetti (2010) investigate the effects of U.S. government spending shock in a dynamic factor model. Unlike the previous two studies, Herrera and Rangaraju (2019) examine the response of state-level per-capita personal income and employment to federal tax changes using news-based implicit tax shocks within the FAVAR framework. They find a significant responsiveness of per-capita GDP and employment to a one percentage point increase in the implicit tax rate, where all states exhibit statistically significant and a similar humped-shaped response, but magnitude and timing are varied across states.

This study follows the macroeconomic approach and employs a FAVAR model but differs in methodology and empirical implementation. Like Herrera and Rangaraju (2019), the focus of this study is on the heterogeneity in the response of state macroeconomic activities to federal tax shocks. While they examine the response of state-level personal income and employment to a news-based implicit tax shock using the FAVAR framework, this study employs narrative tax changes as the exogenous policy shock. Additionally, this study adds a sign restriction with Uhlig's (2005) penalty function in the context of the FAVAR model.

Another study similar to this is Liu and Williams (2019), who investigate the heterogeneous responsiveness of federal tax policy changes at the state level using a small-scale structural VAR model. The findings of their study show that the output and employment responses for more than 50 percent of U.S. states are not statistically significant, which may be related to the limitation of the small-scale SVAR models. In comparison to their study, the state-level FAVAR model of this study employs a broader set of structural characteristics, controlling a broader extent of aggregate and regional variables and allowing more significant results.

## 3. Econometric model, data, and the results

### 3.1 The FAVAR Model

The state-level VAR model used in this study is a FAVAR model with an exogenous tax shock originally developed by Bernanke, Boivin, and Eliasz (2005), and later extended by Stock and Watson (2005, 2016). This model includes four sets of information: a factor from federal level data capturing the comovement of aggregate economic dynamics, a regional factor representing the economic characteristics of the BEA regions, state-specific macroeconomic variable, and narratively identified exogenous tax shocks. The study then estimate a cross-state regression to determine how various state-level characteristics may impact the estimated responses.

Let $F_t \in [f_t^a, f_t^r]$ is the vector of unobserved factors at time $t \ (t = 1, \cdots, T)$ capturing the common comovement at the aggregate and the BEA regional level derived via dynamic factor model. Given a vector of state-level macroeconomic variables $x_t$, a vector of narrative tax shocks $\tau_t$, and $\omega_t$ is a vector that follows standard Gaussian distribution with zero mean and variance-covariance matrix $\Sigma$. The state-level variables are real GDP, personal income, non-farm employment, and consumer price index. We incorporate all the information in the following measurement equation that shows the dynamic relationship between state-specific macroeconomic variable, unobserved factors, and exogenous tax shocks.

$$\begin{aligned} x_t &= a_0 + a_{11}x_{t-1} + a_{12} f_{t-1}^a + a_{13} f_{t-1}^r + b_1\tau_{t-1} + \omega_{1t} \\ f_t^a &= a_1 + a_{21}x_{t-1} + a_{22} f_{t-1}^a + a_{23} f_{t-1}^r + b_2\tau_{t-1} + \omega_{2t} \\ f_t^r &= a_2 + a_{31}x_{t-1} + a_{32} f_{t-1}^a + a_{33} f_{t-1}^r + b_3\tau_{t-1} + \omega_{3t} \end{aligned} \tag{1}$$

We can rewrite the above system of equations in a VAR (1) with the exogenous tax variable $\tau_t$

$$\begin{bmatrix} x_t \\ f_t^a \\ f_t^r \end{bmatrix} = \begin{bmatrix} a_1 \\ a_2 \\ a_3 \end{bmatrix} + \begin{bmatrix} a_{11} & a_{12} & a_{13} \\ a_{21} & a_{22} & a_{23} \\ a_{31} & a_{32} & a_{33} \end{bmatrix} \begin{bmatrix} x_{t-1} \\ f_{t-1}^a \\ f_{t-1}^r \end{bmatrix} + \begin{bmatrix} b_1 \\ b_2 \\ b_3 \end{bmatrix} \tau_{t-1} + \begin{bmatrix} \omega_{1t} \\ \omega_{2t} \\ \omega_{3t} \end{bmatrix} \tag{2}$$

Let $y_t \in \{x_t, f_{a,t}, f_{r,t}\}$, then the general representation of the above system for $p$ and $q$ lags can be written as follows:

$$y_t = A_0 + A_1 y_{t-1} + \cdots + A_p y_{t-p} + B_1\tau_{t-1} + \cdots + B_q\tau_{t-q} + \omega_t \tag{3}$$

where A and B's are coefficient matrix; $\omega_t$ follows standard multivariate normality assumptions. If we assume an auto-regression process for the exogenous tax shock $\tau_t$ itself:

$$\tau_t = C_0 + C_1\tau_{t-1} + \cdots C_r\tau_{t-r} + \upsilon_t \tag{4}$$

Here $E(\upsilon_t \mid y_{t-1}, \tau_{t-1}) = 0$, that implies that $y_t$ does not Granger Cause $\tau_t$ (as shown in Alam, 2021a) and the weak form exogeneity in the above system. In general, one can also consider a contemporary $\tau_t$ in equation (3), then the equation can be written as:

$$y_t = A_0 + A_1 y_{t-1} + \cdots + A_p y_{t-p} + B_0\,\tau_t + B_1\tau_{t-1} + \cdots + B_q\tau_{t-q} + \omega_t \tag{5}$$

If $B_0$ is not equal to zero, then it may be possible that $y_t$ has an indirect influence on $\tau_t$ through the independence between the two error terms. In that case, equations (4) and (5) together form a system of simultaneous equations. In reduced form, it is as follows:

$$\begin{aligned} y_t = A_0 + B_0 C_0 + A_1 y_{t-1} + \cdots + A_p y_{t-p} + \\ B_0 (C_1 \tau_{t-1} + \cdots C_r \tau_{t-r}) + B_1 \tau_{t-1} + \cdots + B_q \tau_{t-q} + \omega_t + B_0 \upsilon_t \end{aligned} \qquad (6)$$

Equation (6) is another form of equation (3) and implies that the absence of contemporaneous impact of $\tau_t$ on $y_t$ in equation (3) is not a loss of information.

Now considering all the coefficient matrices, A, B, and C, and imposing exogenity restrictions between $y_t$ and $\tau_t$ , and assume $p = q = r$ , we can represent the equation (3) and (4) as a VAR (p) model:

$$\begin{bmatrix} y_t \\ \tau_t \end{bmatrix} = \begin{bmatrix} A_0 \\ C_0 \end{bmatrix} + \begin{bmatrix} A_1 & B_1 \\ C_1 & C_2 \end{bmatrix} \begin{bmatrix} y_{t-1} \\ \tau_{t-1} \end{bmatrix} + \cdots + \begin{bmatrix} A_p & B_p \\ C_{3p} & C_{4p} \end{bmatrix} \begin{bmatrix} y_{t-p} \\ \tau_{t-p} \end{bmatrix} + \begin{bmatrix} \omega_t \\ \upsilon_t \end{bmatrix} \qquad (7)$$

Now, the Granger causality condition implies the exogenity restriction of zero in the coefficient matrix in equation (7), that is $C_1 = C_{3p} = 0$ ,

$$\begin{bmatrix} y_t \\ \tau_t \end{bmatrix} = \begin{bmatrix} A_0 \\ C_0 \end{bmatrix} + \begin{bmatrix} A_1 & B_1 \\ O & C_2 \end{bmatrix} \begin{bmatrix} y_{t-1} \\ \tau_{t-1} \end{bmatrix} + \cdots + \begin{bmatrix} A_p & B_p \\ O & C_{4p} \end{bmatrix} \begin{bmatrix} y_{t-p} \\ \tau_{t-p} \end{bmatrix} + \begin{bmatrix} \omega_t \\ \upsilon_t \end{bmatrix} \qquad (8)$$

Additionally, if we consider no lag for $\tau_t$ (contemporaneous impact of $\tau_t$ on $y_t$ ), then $C_2$ and $C_{4p}$ will be the identity matrix. Putting these two restrictions together, equation (7) for VAR (1) can be written as:

$$\begin{bmatrix} y_t \\ \tau_t \end{bmatrix} = \begin{bmatrix} A_0 \\ C_0 \end{bmatrix} + \begin{bmatrix} A_1 & B_1 \\ O & I \end{bmatrix} \begin{bmatrix} y_{t-1} \\ \tau_t \end{bmatrix} + \begin{bmatrix} \omega_t \\ O \end{bmatrix} \qquad (9)$$

So, the weak form of exogeneity under the Granger-causality in equation (8) and (9) implies that $y_t$ does not Granger-cause $\tau_t$ . We maintain standard assumption in VAR equations is that the errors are *i.i.d* and normally distributed. So,

$$E[\omega_t \mid \{y_{t-j}\}_{j=1}^{\infty}, \{\tau_{t-i}\}_{i=1}^{\infty}] = 0 \ \ (\in R^k)$$

$$E[\upsilon_t \mid \{y_{t-j}\}_{j=1}^{\infty}, \{\tau_{t-i}\}_{i=1}^{\infty}] = 0 \ \ (\in R^m)$$

$$\begin{pmatrix} \omega_t \\ \upsilon_t \end{pmatrix} \sim i.i.d.\ N_{k+m}\left[\begin{pmatrix} 0 \\ 0 \end{pmatrix}, \begin{pmatrix} \Sigma_{11} & \Sigma_{12} \\ \Sigma_{21} & \Sigma_{22} \end{pmatrix}\right] \qquad (10)$$

Under the assumption of weak exogeneity and strong exogeneity, that $\Sigma_{21} = 0$ and $\Sigma_{12} = 0$ satisfy, then the innovation response of $y_t$ to a unit shock in the innovation of $\tau_t$ can be estimated from the structural VAR (1) model in equation (9). The objective in the subsequent empirical analysis is to estimate the response of state-level macroeconomic variable ($x_t$) up to 10-year forecasting horizons to a one percent cut of federal tax ($\tau_t$).

### 3.2 Data

This study uses annual data from 1977 to 2018. The dataset for the aggregate factor ($f^a$) consists of a large panel of 132 macroeconomic series (Alam, 2021a, 2021b). The series includes federal level information on real output, components of real activity, price levels, housing market, industrial production, labor market, money and financial market. The dataset for the BEA regional factor ($f^r$) consists of 48 series. I consider six macroeconomic series for each of the BEA region[2]: per capita personal income, personal consumption expenditures, civilian unemployment rate, and total deposits in commercial banks, consumer price index, and the number of new private housing units.

Annual frequency of narrative tax shocks is a subsample of extended narrative dataset developed Alam (2021a), which includes 13 personal income tax changes and 11 corporate income tax changes for 1977-2018 periods. A more detailed description of data sources, the complete list of federal and regional dataset, and a list of U.S. states used in this study are in Appendix I - IV. The state-level macroeconomic variables used in the VAR model include real GDP, personal income, consumer price index, unemployment rate, and non-farm employment. From 1977 to 2018, the annual frequency of these series comes from BEA regional accounts and the St. Louis Federal Reserve. All series are seasonally adjusted, and nominal GDP and personal income are transformed into the real series using the GDP deflator. Summary statistics for the yearly sample of 1st, median, and 50th states between 1977 and 2018 are shown in Table 1 along with their correlation with the U.S. aggregates.

[2] The selection of these variables is based on the data availability for the BEA regions and the list of key macroeconomic, credit and housing market variables considered in Belviso and Milani (2005) and Stock and Watson (2016).

### 3.3 Factor and model estimation

I estimate common factors from the panel of aggregate and BEA regional macroeconomic series using a dynamic factor model. The regional factor ( $f^r$ ) is from the eight BEA regional economic accounts and the aggregate factor ( $f^a$ ) is from a large panel of federal macroeconomic series. Figure 1 and 2 plot the aggregate and regional factor series and their relationship with two leading U.S. macroeconomic activity indices, the Chicago Fed National Activity Index (CFNAI) and the Brave-Butters-Kelley's (Brave, Butters, and Kelley, 2019) Coincident Index. The factor accurately tracks the growth and dynamics of U.S. aggregate and regional economic activity. With the estimated factors, narrative tax changes, and state-level variables, the estimation of equation (1) uses sign restrictions with Uhlig's (2005) penalty function.

Table 1: Summary statistics of state-level macroeconomic variables

Summary statistics of state-level macroeconomic variables

| Variable | States | Mean | Std.dev | Max | Min | Corr with U.S. aggregates |
|---|---|---|---|---|---|---|
| Real GDP growth | **U.S.** | **2.6** | **1.85** | **7.13** | **-2.54** | **1** |
| | 1st (CA) | 3.31 | 2.6 | 8.06 | -4.01 | 0.82 |
| | Median (AL) | 2.17 | 2.26 | 6.42 | -3.93 | 0.81 |
| | 50th (VT) | 2.88 | 2.80 | 10.00 | -2.71 | 0.67 |
| Disp. Pers. Inc | **U.S.** | **5.08** | **1.81** | **8.82** | **-0.30** | **1** |
| | 1st (CA) | 5.38 | 2.17 | 9.69 | -0.15 | 0.88 |
| | Median (AL) | 4.77 | 1.89 | 8.41 | 0.25 | 0.83 |
| | 50th (WY) | 4.99 | 3.76 | 8.43 | -4.22 | 0.43 |
| Employment growth | **U.S.** | **1.63** | **1.41** | **4.42** | **-3.11** | **1** |
| | 1st (CA) | 1.98 | 2.06 | 6.27 | -3.95 | 0.89 |
| | Median (KY) | 1.25 | 1.50 | 4.16 | -3.22 | 0.87 |
| | 50th (WY) | 1.56 | 2.74 | 8.34 | -4.66 | 0.39 |
| Inflation (% change in CPI) | **U.S.** | **2.92** | **1.51** | **7.63** | **-0.32** | **1** |
| | 1st (CA) | 3.42 | 1.9 | 10.03 | -0.02 | 0.74 |
| | Median (CO) | 3.20 | 1.9 | 9.39 | -0.65 | 0.74 |
| | 50th (AK) | 2.81 | 1.3 | 5.94 | -0.48 | 0.95 |

The advantage of the penalty function is that it implements the unconstrained optimization by quadratic approximation (Powell, 2002). The model estimation algorithm employs the following steps with a given set of sign restrictions in Table 2:

(1) Estimate the unrestricted VAR in equation (9) to obtain coefficient matrix ( $\hat{\Lambda}^F$ , $\hat{\Lambda}^T$ ) and the variance-covariance matrix $\hat{\Sigma}_{\omega}$ .

(2) Apply a Cholesky decomposition to the model and extract the orthogonal innovations.

(3) Estimate impulse responses using the orthogonal innovations from step 2.

(4) Given $\hat{\Lambda}^F$ , $\hat{\Lambda}^T$ , and $\hat{\Sigma}_{\omega}$ , a random impulse vector is drawn following Uhlig's (2005) penalty function approach.

(5) Impulse responses in step-3 are multiplied by the randomly drawn impulse vector from step-4 and examine if the prior sign restrictions are satisfied by minimizing the penalty function.

(6) If *a priori* sign restrictions hold, the model keeps the impulse responses; otherwise, the system drops the draw and return to step 2.

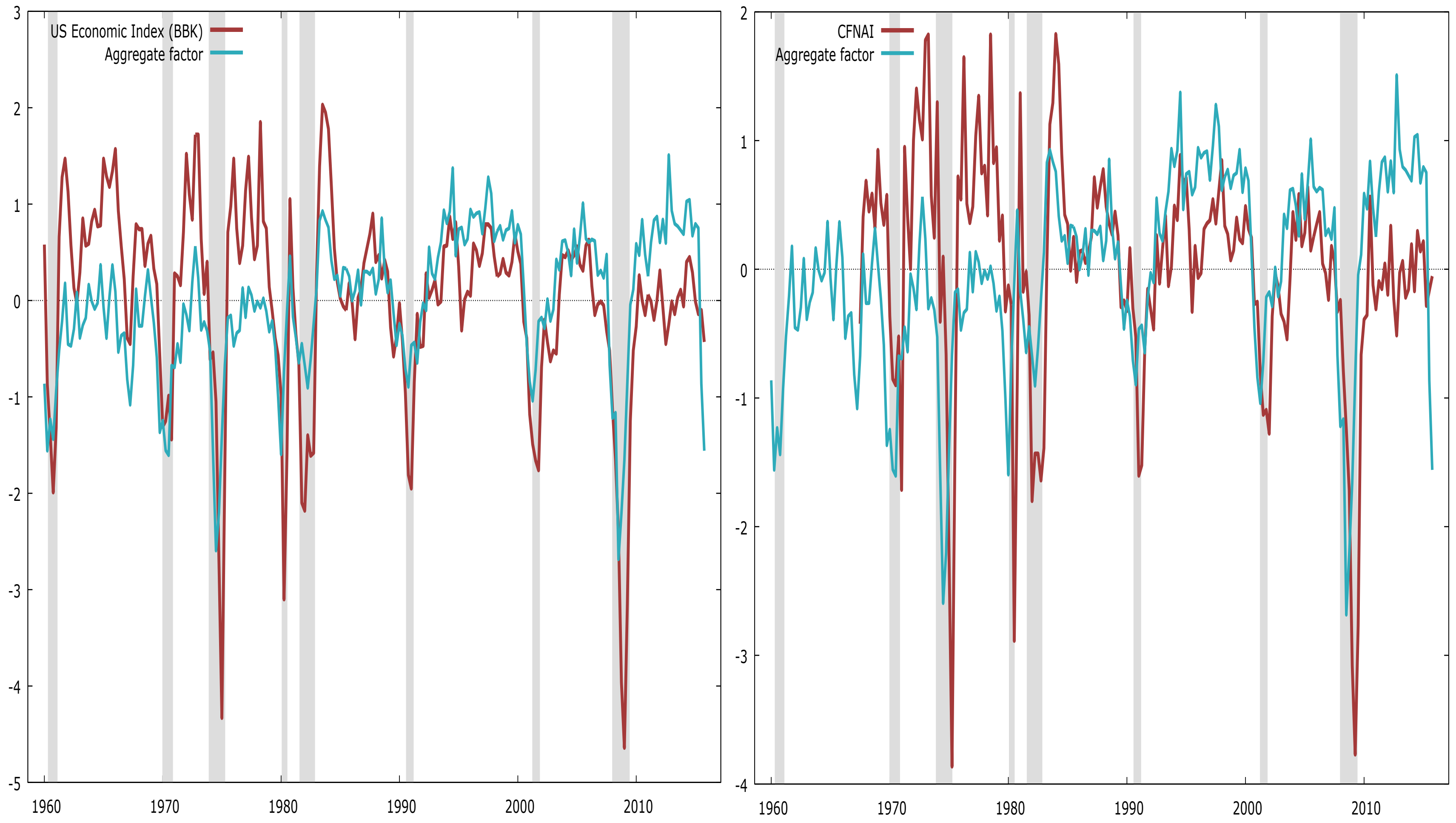

Figure 1: Comovement between aggregate factor and the CFNAI index. The shaded vertical bars indicate official periods of economic recession for the U.S., as announced by the National Bureau of Economic Research.

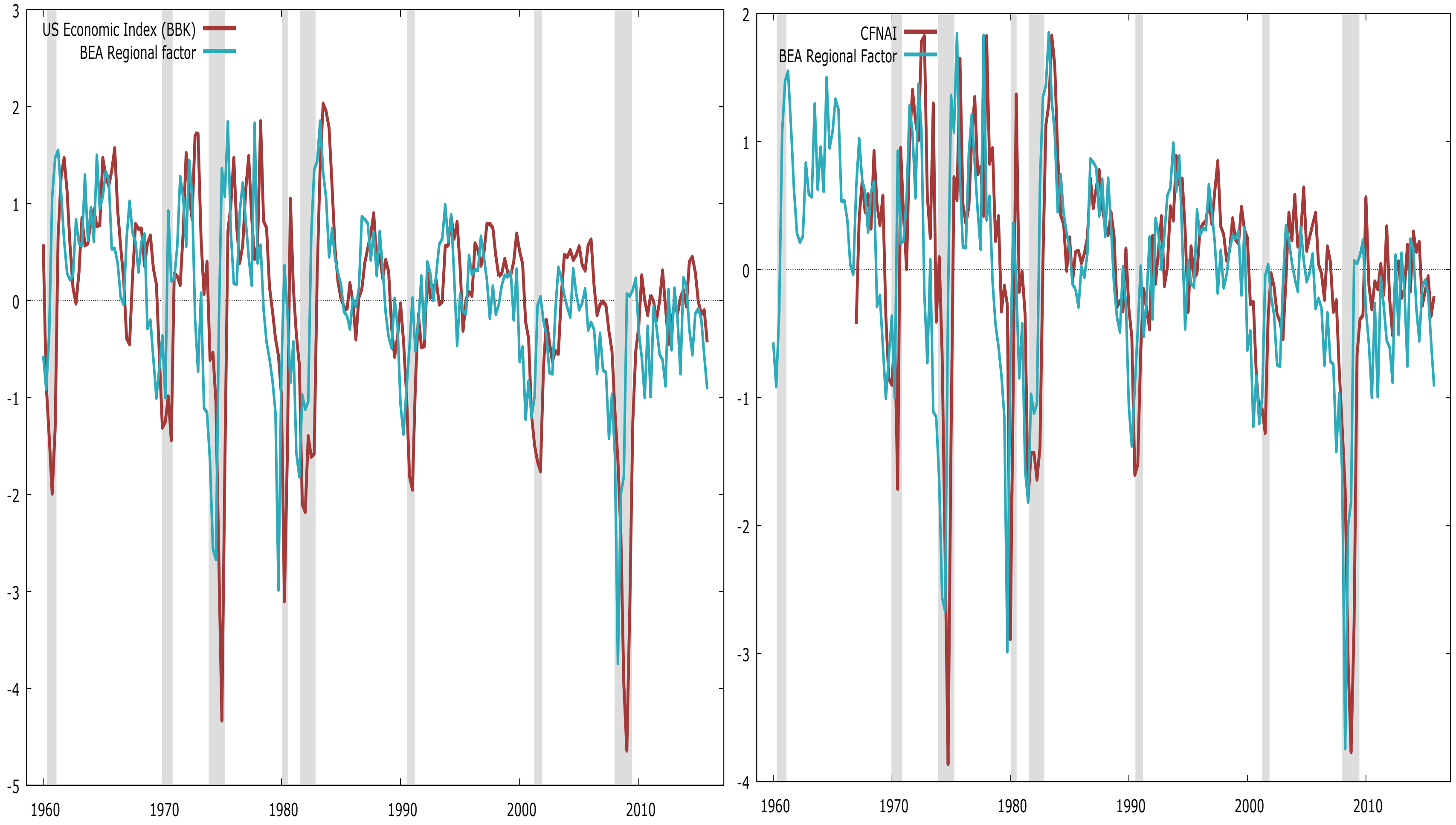

Figure 2: Comovement between the BEA factor and the CFNAI index. The shaded vertical bars indicate official periods of economic recession for the U.S., as announced by the National Bureau of Economic Research.

Finally, I evaluate the significance of estimated impulse responses and the identification of shocks using the Median-Target (MT) method suggested by Fry and Pagan (2011). The method identifies how significant the FAVAR parameters and how close the estimated impulse responses to the median impulse responses.

Table 2: Sign restriction on the variables of the FAVAR model

| | Sign restriction | | | | | | | |
|---|---|---|---|---|---|---|---|---|
| Variable | GDP | DPI | CPI | EMP | PIT | CIT | $F^a$ | $F^r$ |
| Personal income tax cut shocks | + | + | + | + | ≅0 | ≅0 | ≅0 | ≅0 |
| Corporate income tax cut shocks | + | + | + | + | ≅0 | ≅0 | ≅0 | ≅0 |

≅0 indicates no restriction.

### 3.4 Main results: response of macroeconomic variables

For each state, the model estimates the response of GDP, disposable personal income, employment, and price level to a one percent cut in federal personal and corporate income tax. The cumulative impulse responses of GDP, employment, personal income, and CPI are summarized in Figures 3 and 4 over a 10-year horizon. The response values are normalized into a range of [0, 1]. The visualization of cumulative impulse responses depends on the classification scheme that divides the states into four categories. States with deep and light blue colors denote the responses within the range of [1, 0.35], while grey-white and white colors represent the range of [0.34, 0] for less responsive states. As shown in both figures, the responses are not homogenous across the U.S. states. The states are also ranked based on their cumulative responses over a 10-year horizon. Figures 5 and 6 show the impulse response function of GDP, personal income, employment, and CPI for 1st, 2nd, 25th, and 50th states. The area in between two blue lines shows the 95% bootstrap confidence intervals. The impulse response function of all 50 states and states level are in Appendix V (Figure 13-20).

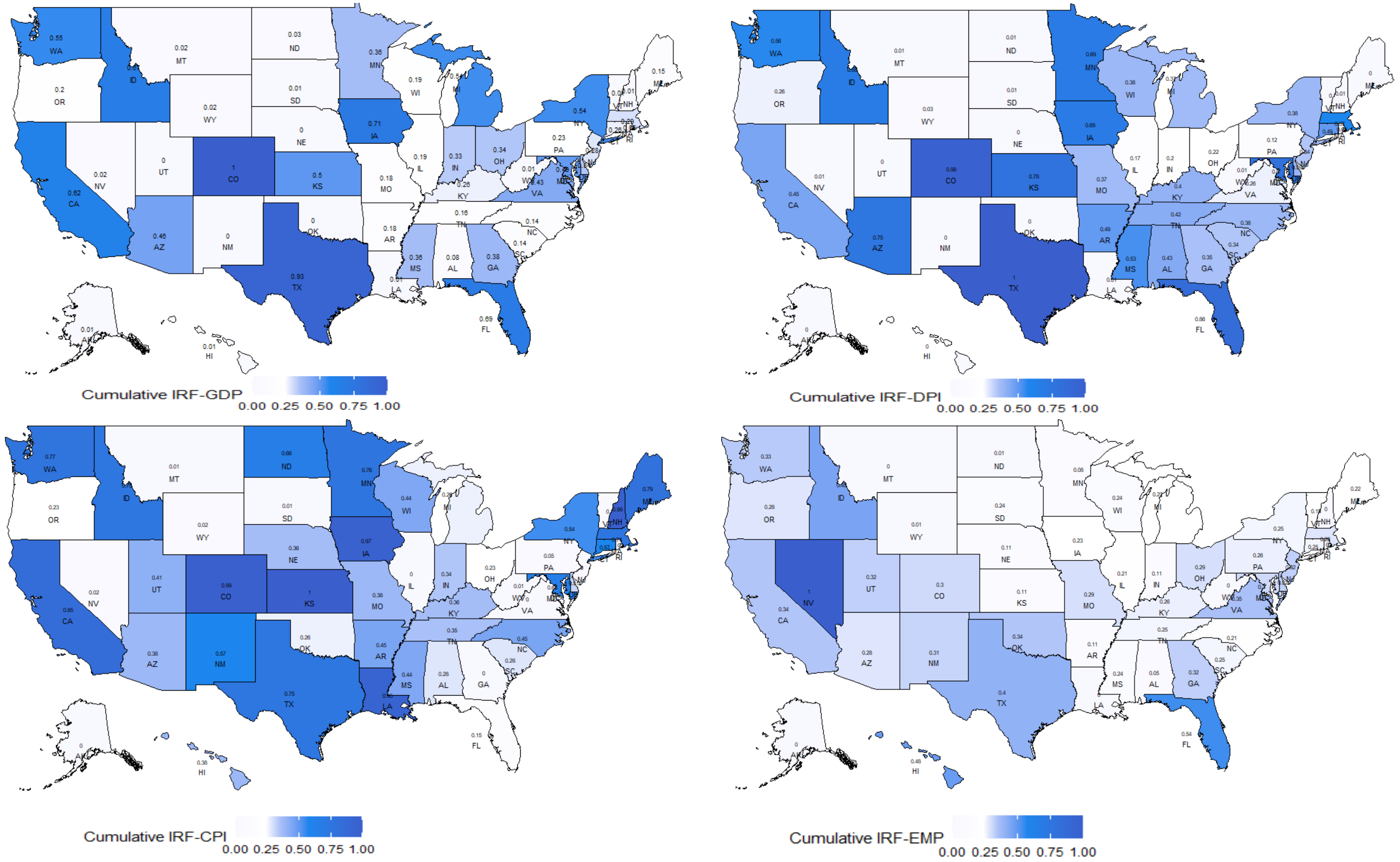

Figure 3. Cumulative responses of states (PIT).

Horizontal bar under the map presents the range of the magnitude of the normalized values of cumulative responses.

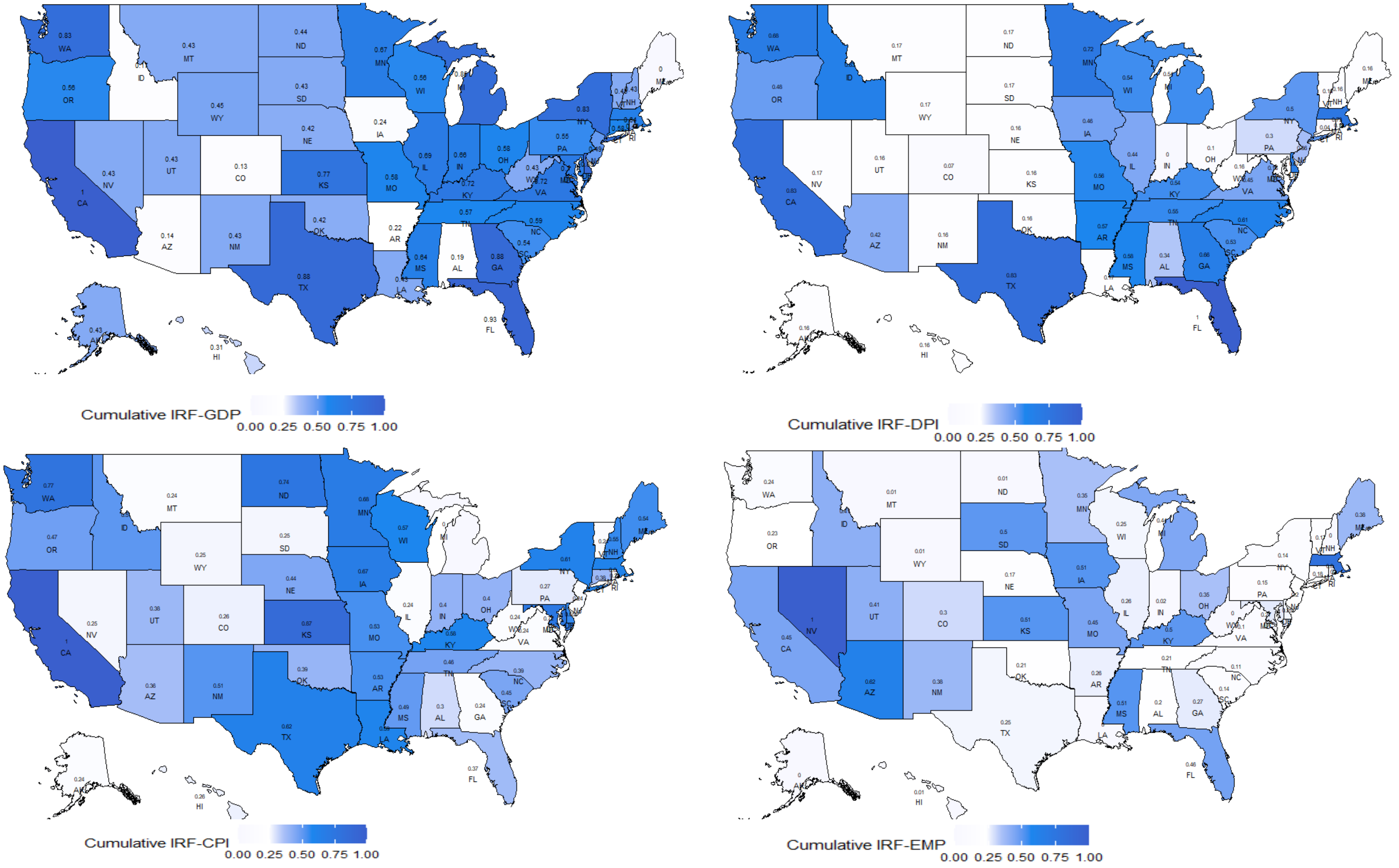

Figure 4. Cumulative responses of states (CIT).
Horizontal bar under the map presents the range of the magnitude of the normalized values of cumulative responses.

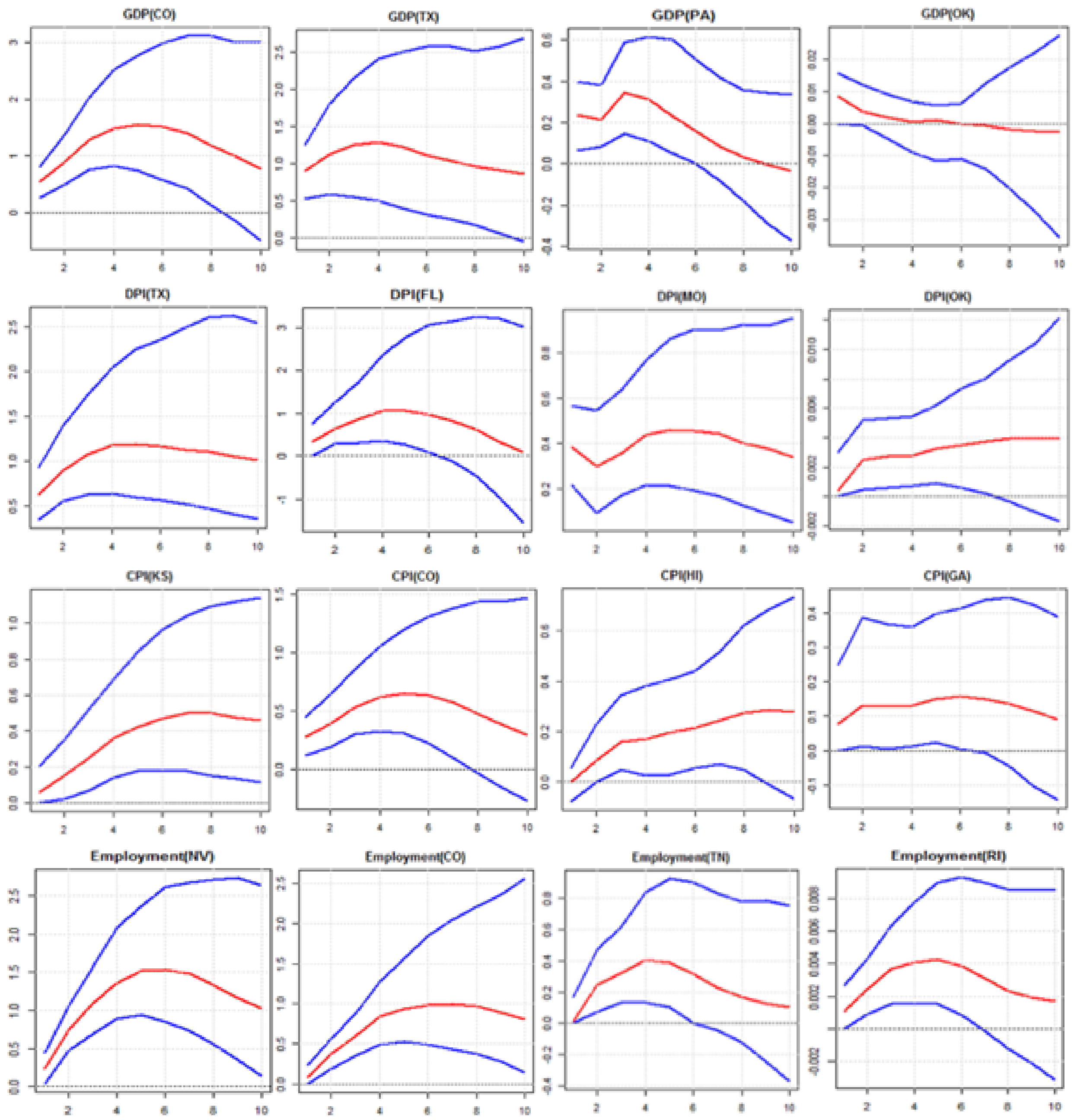

Figure 5: Impulses responses (PIT). IRFs of GDP, DPI, CPI and employment to a cut in personal income tax rate. The responses of 1st, 2nd, 25th and 50th states are presented here where states are ordered by their cumulative impulse responses over a 10-year horizon.

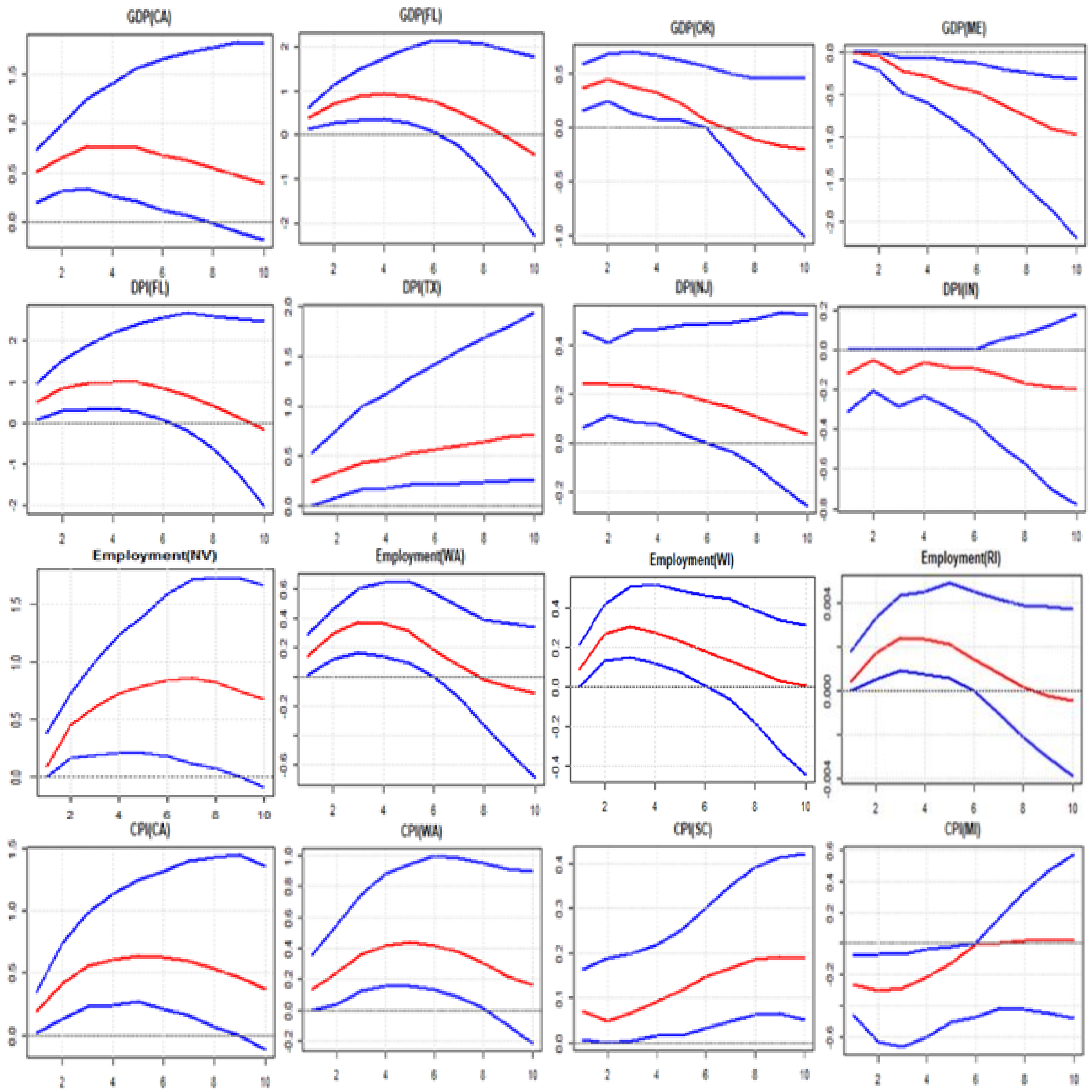

Figure 6: Impulses responses (CIT). IRFs of GDP, DPI, CPI and employment to a cut in corporate income tax rate. The responses of $1^{st}$, $2^{nd}$, $25^{th}$ and $50^{th}$ states are presented here where states are ordered by their cumulative impulse responses over a 10-year horizon.

In the responses of real GDP and disposable personal income, the state economies do react heterogeneously in magnitude and directions, and, in most states, these reactions are statistically significant. A one percent cut in personal income tax leads to a rise in GDP for about thirty-three states. The personal income rises in thirty-nine states while decreasing in GDP in fourteen states and having insignificant effects in the remaining three states. The highest response of real GDP in the first ten states lies within the range from 0.8 percent to 2.2 percent in Texas, Colorado, California, Iowa, Washington, and New York. Like the response of personal income tax cut, thirty-seven states exhibit a positive response in real GDP to a similar cut in federal corporate income tax, with the maximum rise estimated at 1.2 percent. The magnitude of the rise is largest in Florida, Texas, Georgia, California, and Michigan, with real GDP rising by about 0.75 percent to 1.2 percent, which is, on average, slightly lower than the magnitude of response to a personal income tax cut.

The expansionary effects on personal income from a corporate income tax cut are also observed for about forty states, with the highest responses range from 0.3 percent to 1.2 percent. Personal income decreases for six states, while four states show insignificant responses. States with the insignificant and lowest response (from -5.4 to -0.015 percent) in real GDP to tax changes include Maine, Utah, Nevada, New Mexico, Alabama, Idaho, and Oklahoma.

One of the primary objectives of federal tax cut is to accelerate nonfarm private, manufacturing, and business sector job growth while federal taxes are only one part of a state economic picture that includes labor costs. In addition to other economic factors, state's fiscal structures also play an outsized role in state's employment growth. Empirical studies, as in Moretti and Wilson (2017), and Serrato and Zidar (2018) show that job-creating investment flows from high-tax states to low-tax states. This study measures employment as the number of newly added jobs at the state's nonfarm business, manufacturing, and private industries. Employment rises in all states, and tax cuts affect states heterogeneously at the 10-year horizons. The highest response to a personal income tax cut in employment lies within the range from 0.6 percent to 1.5 percent in Colorado, Florida, Texas, Arizona, Idaho, South Dakota, Virginia, and Washington.

The magnitudes are also heterogeneous across all states, while states with no personal and corporate income taxes or moderate state tax rates show relatively a higher response to either tax changes. States with lower income taxes and states with more financial, manufacturing, and service sectors in their GDP shares, such as Utah, Delaware, South Dakota, Washington, Florida, and Texas exhibit a higher employment growth expansion. The employment responses to a similar corporate tax cut are relatively lower, where the magnitude lies within the range from 0.2 percent to 0.9 percent. So, a cut in personal income tax produces a higher and prolonged impact on employment growth than the cut in corporate income taxes. These results are relatively similar to those of Romer and Romer (2009) and Marten and Ravn (2013), particularly in the direction of the impulse response functions. However, the magnitude and the persistence of the impulse response functions appear to be quite different for most states.

Consistent with the prediction of standard Keynesian models, a cut in both taxes significantly increase the consumer price index for almost all states except for Delaware and South Carolina. The magnitude of the cumulative response of the price level to a personal income tax cut range between 0.02 and 5.8 percent. The highest responsive states are California, Washington, Kansas, Colorado, Minnesota, New Hampshire, and New York, where the cumulative responses are on average lies between 2.30 percent and 5.8 percent. In contrast, a few states like Alabama, Montana, Virginia, Arkansas, and Georgia are less responsive, with maximum cumulative impulse responses accounted for 0.96 percent at the 10-year horizon.

In addition to cumulative impulse response functions, the model also estimates the tax shocks' contribution to the variance decomposition of GDP, personal income, employment, and price levels. Tables 3 and 4 show substantial heterogeneity in the percentage of variance decomposition. Taking the proportion of GDP variation as an example, personal income tax shocks explain the maximum amount of variation, about 58 percent of GDP for Nevada, while the variation is only 2.32 percent for North Carolina. Personal income tax shocks also explain a significant amount of variation for employment and price levels, on average, from 17 percent up to 27 percent for Texas, Michigan, Nevada, Minnesota, Utah, Georgia, and South Dakota. Relative to personal income tax shocks, corporate income tax shocks, on average, accounted for a smaller portion of the variance. As shown in Tables 3 and 4, above the mean amount of shocks' variation for both real GDP and personal income lies in between 18 percent and 37 percent for Washington, South Dakota, Nevada, Florida, Delaware, and California.

Table 3: Forecast Error Variance Decomposition (Real GDP)

| | PIT Shock Horizon (in year) | | | CIT Shock Horizon (in year) | | |
|---|---|---|---|---|---|---|
| State | 1 | 5 | 10 | 1 | 5 | 10 |
| AK | 0.52 | 6.44 | 8.30 | 1.08 | 4.63 | 5.69 |
| AL | 12.45 | 13.45 | 14.11 | 32.63 | 16.81 | 12.71 |
| AR | 22.93 | 17.59 | 16.10 | 1.45 | 7.08 | 9.96 |
| AZ | 54.64 | 31.97 | 22.88 | 7.81 | 10.39 | 10.28 |
| CA | 43.10 | 29.16 | 23.19 | 37.50 | 23.35 | 16.67 |
| CO | 31.91 | 25.55 | 22.44 | 10.83 | 11.24 | 11.03 |
| CT | 20.57 | 16.02 | 14.59 | 19.31 | 15.16 | 14.48 |
| DE | 19.11 | 13.22 | 9.71 | 36.83 | 20.77 | 14.37 |
| FL | 29.16 | 24.12 | 20.10 | 26.86 | 22.87 | 16.53 |
| GA | 21.92 | 18.91 | 16.29 | 16.14 | 16.98 | 14.71 |
| HI | 3.31 | 6.50 | 6.98 | 4.63 | 20.29 | 29.55 |
| IA | 25.53 | 20.08 | 20.12 | 19.25 | 20.51 | 21.33 |
| ID | 39.64 | 28.84 | 24.33 | 6.81 | 10.11 | 11.48 |
| IL | 17.14 | 14.70 | 14.04 | 15.53 | 15.80 | 14.14 |
| IN | 36.66 | 21.10 | 18.30 | 14.15 | 12.53 | 11.77 |
| KS | 17.54 | 17.75 | 19.07 | 6.03 | 12.44 | 12.57 |
| KY | 12.52 | 13.33 | 13.48 | 5.68 | 9.84 | 12.87 |
| LA | 5.51 | 15.35 | 17.21 | 0.26 | 7.53 | 8.33 |
| MA | 23.02 | 19.21 | 15.46 | 9.28 | 13.41 | 15.88 |
| MD | 29.73 | 24.47 | 22.44 | 19.39 | 17.84 | 16.88 |
| ME | 33.55 | 25.41 | 16.82 | 14.52 | 14.17 | 14.34 |
| MI | 27.37 | 21.00 | 18.20 | 11.99 | 12.58 | 13.17 |
| MN | 39.09 | 26.80 | 22.62 | 31.17 | 20.66 | 18.76 |
| MO | 21.02 | 18.62 | 17.76 | 13.37 | 13.29 | 14.34 |
| MS | 18.10 | 16.90 | 15.58 | 12.32 | 12.63 | 11.54 |
| MT | 44.92 | 31.56 | 26.81 | 12.11 | 13.50 | 14.36 |
| NC | 2.22 | 11.85 | 11.89 | 2.27 | 10.99 | 10.77 |
| ND | 28.07 | 23.90 | 22.38 | 28.48 | 21.33 | 17.61 |
| NE | 14.71 | 11.98 | 13.03 | 14.27 | 11.82 | 11.63 |
| NH | 18.87 | 19.08 | 17.72 | 8.41 | 9.75 | 11.71 |
| NJ | 2.67 | 7.10 | 8.85 | 2.45 | 6.29 | 9.20 |
| NM | 4.26 | 9.11 | 10.33 | 2.80 | 7.13 | 9.01 |
| NV | 57.93 | 38.06 | 26.12 | 26.31 | 20.25 | 14.79 |
| NY | 21.57 | 18.91 | 15.86 | 6.39 | 10.71 | 12.00 |
| OH | 24.06 | 18.13 | 16.27 | 12.47 | 12.71 | 12.15 |
| OK | 5.93 | 6.10 | 6.82 | 3.21 | 8.18 | 9.42 |
| OR | 20.92 | 17.59 | 18.81 | 13.91 | 13.80 | 15.77 |
| PA | 19.58 | 17.97 | 16.81 | 24.76 | 17.59 | 16.33 |
| RI | 30.30 | 22.22 | 17.16 | 25.94 | 20.25 | 15.53 |
| SC | 23.40 | 18.95 | 16.98 | 17.17 | 15.12 | 15.28 |
| SD | 42.24 | 26.48 | 19.71 | 23.99 | 20.96 | 20.96 |
| TN | 21.82 | 18.06 | 15.25 | 19.11 | 16.35 | 15.36 |
| TX | 49.62 | 30.55 | 23.91 | 2.06 | 7.77 | 10.09 |
| UT | 1.28 | 6.80 | 10.69 | 1.01 | 4.96 | 7.06 |
| VA | 18.11 | 16.30 | 15.58 | 21.25 | 17.82 | 17.17 |
| VT | 36.95 | 28.48 | 18.28 | 35.71 | 25.91 | 20.66 |
| WA | 26.18 | 21.97 | 19.58 | 19.64 | 14.40 | 13.08 |
| WI | 42.07 | 24.84 | 19.93 | 30.38 | 17.79 | 15.33 |
| WV | 10.42 | 19.96 | 21.13 | 3.80 | 11.05 | 12.25 |
| WY | 29.11 | 25.71 | 26.31 | 6.15 | 15.04 | 17.52 |

Table 4: Forecast Error Variance Decomposition (DPI)

| State | PIT Shock Horizon (in year) 1 | 5 | 10 | CIT Shock Horizon (in year) 1 | 5 | 10 |
|---|---|---|---|---|---|---|
| AK | 30.33 | 9.45 | 9.16 | 8.75 | 5.87 | 6.10 |
| AL | 7.65 | 13.48 | 14.11 | 10.03 | 10.12 | 9.87 |
| AR | 27.13 | 22.63 | 18.81 | 33.41 | 20.29 | 16.95 |
| AZ | 30.75 | 23.67 | 18.07 | 24.87 | 16.23 | 12.72 |
| CA | 13.43 | 17.38 | 16.43 | 28.35 | 25.19 | 21.07 |
| CO | 15.55 | 19.00 | 18.13 | 3.33 | 7.99 | 11.03 |
| CT | 4.32 | 11.09 | 13.01 | 5.70 | 11.91 | 13.69 |
| DE | 8.60 | 11.73 | 11.36 | 15.45 | 17.80 | 15.21 |
| FL | 15.53 | 19.53 | 18.59 | 17.30 | 18.83 | 15.97 |
| GA | 16.55 | 18.98 | 16.91 | 8.79 | 11.35 | 10.99 |
| HI | 22.73 | 18.32 | 14.66 | 33.56 | 28.18 | 21.89 |
| IA | 11.82 | 16.46 | 21.25 | 9.07 | 10.52 | 11.77 |
| ID | 24.54 | 24.22 | 20.16 | 26.93 | 18.28 | 15.88 |
| IL | 9.25 | 13.71 | 13.60 | 7.91 | 13.95 | 13.39 |
| IN | 13.59 | 16.00 | 17.05 | 7.98 | 9.49 | 11.59 |
| KS | 13.08 | 15.54 | 20.93 | 15.49 | 21.22 | 19.00 |
| KY | 9.28 | 14.83 | 18.27 | 9.56 | 12.55 | 12.23 |
| LA | 24.12 | 21.19 | 20.18 | 14.04 | 16.67 | 17.01 |
| MA | 10.94 | 14.60 | 14.94 | 11.81 | 13.55 | 14.10 |
| MD | 25.73 | 28.49 | 26.27 | 19.07 | 21.02 | 19.74 |
| ME | 21.20 | 24.01 | 19.94 | 19.95 | 12.41 | 13.94 |
| MI | 24.24 | 20.88 | 17.89 | 9.53 | 12.49 | 12.62 |
| MN | 24.93 | 24.90 | 24.54 | 11.59 | 14.45 | 15.51 |
| MO | 26.04 | 22.79 | 22.11 | 8.55 | 12.36 | 14.34 |
| MS | 30.98 | 23.80 | 21.04 | 20.36 | 16.88 | 15.01 |
| MT | 31.52 | 30.58 | 26.57 | 13.95 | 14.05 | 14.51 |
| NC | 3.93 | 12.05 | 12.23 | 10.33 | 14.32 | 13.07 |
| ND | 22.17 | 22.00 | 21.06 | 23.51 | 16.82 | 13.60 |
| NE | 6.14 | 12.41 | 13.94 | 3.33 | 11.28 | 12.84 |
| NH | 22.43 | 22.72 | 19.58 | 6.08 | 8.95 | 11.17 |
| NJ | 7.23 | 9.31 | 9.51 | 5.17 | 7.91 | 8.68 |
| NM | 29.55 | 24.04 | 21.74 | 8.27 | 12.95 | 12.07 |
| NV | 24.80 | 36.47 | 26.79 | 19.27 | 20.02 | 15.09 |
| NY | 10.95 | 15.18 | 15.15 | 5.79 | 10.30 | 10.79 |
| OH | 8.67 | 12.53 | 12.66 | 2.34 | 5.69 | 6.88 |
| OK | 0.74 | 6.52 | 7.44 | 2.01 | 7.39 | 8.09 |
| OR | 11.52 | 14.45 | 16.13 | 10.60 | 13.23 | 14.38 |
| PA | 3.77 | 10.58 | 13.82 | 2.89 | 11.69 | 13.69 |
| RI | 6.16 | 14.97 | 15.59 | 5.11 | 13.43 | 13.82 |
| SC | 16.26 | 15.59 | 15.26 | 15.20 | 14.10 | 14.93 |
| SD | 24.43 | 20.57 | 18.52 | 25.72 | 22.92 | 21.41 |
| TN | 27.78 | 22.19 | 19.53 | 19.10 | 16.95 | 15.97 |
| TX | 26.31 | 31.36 | 27.08 | 6.00 | 7.81 | 9.35 |
| UT | 25.36 | 18.13 | 18.16 | 14.97 | 13.03 | 13.17 |
| VA | 23.53 | 22.42 | 19.48 | 7.83 | 11.86 | 14.38 |
| VT | 32.49 | 27.43 | 18.48 | 28.68 | 24.00 | 16.50 |
| WA | 16.27 | 18.22 | 17.23 | 19.33 | 18.12 | 16.85 |
| WI | 13.80 | 19.00 | 18.52 | 9.88 | 14.97 | 14.84 |
| WV | 28.02 | 23.48 | 22.35 | 11.03 | 11.55 | 11.99 |
| WY | 29.36 | 28.22 | 27.58 | 7.92 | 10.31 | 11.20 |

### 3.5 Reliability of the FAVAR estimates

The reliability of the identification of the FAVAR estimates is examined using the Median-Target (M-T) approach proposed by Fry and Pagan (2011). The main advantage of the M-T method (Furlanetto, Ravazzolo, and Sarferaz, 2019; Mangadi and Sheen, 2017) is that it examines the significance of the estimated impulse responses by minimizing the standardized differences between the impulse responses from the FAVAR model and the responses drawn from the M-T method. The M-T method estimates a single impulse response function by optimizing the gap between the impulse responses of two models and finally provides a unique impulse response function that is the closest possible one to the response of the FAVAR model.

Figures 7-10 show the impulse responses of the M-T method (solid red lines) and compares them with the impulse responses of the FAVAR model (dashed blue lines). As in the benchmark model, the narrative tax shock is identified by imposing sign restrictions on the macroeconomic variables with Uhlig's (2005) penalty function. The M-T method replicates the impulse response functions of real GDP, personal income, CPI, and employment. For each impulse response function of these four variables, responses are ranked by state's cumulative responses over a ten-year horizon, and figure 7-10 reports the responses for the 1st, 2nd, 25th, and 50th states. The method compares how the impulse responses change once the M-T method identifies the single best median impulse vector. The best possible responses for all macroeconomic variables match the FAVAR models' responses perfectly on impact and a 10-year horizon. The method also delivers 95 percent confidence intervals. The confidence intervals associated with the FAVAR impulse responses and M-T's responses are statistically significant and do not contain zero for, except for GDP responses in Oklahoma. So, Fry and Pagan's (2011) M-T diagnostic tool validates the FAVAR model's specification and the identification of the tax shocks in the estimation of the benchmark model.

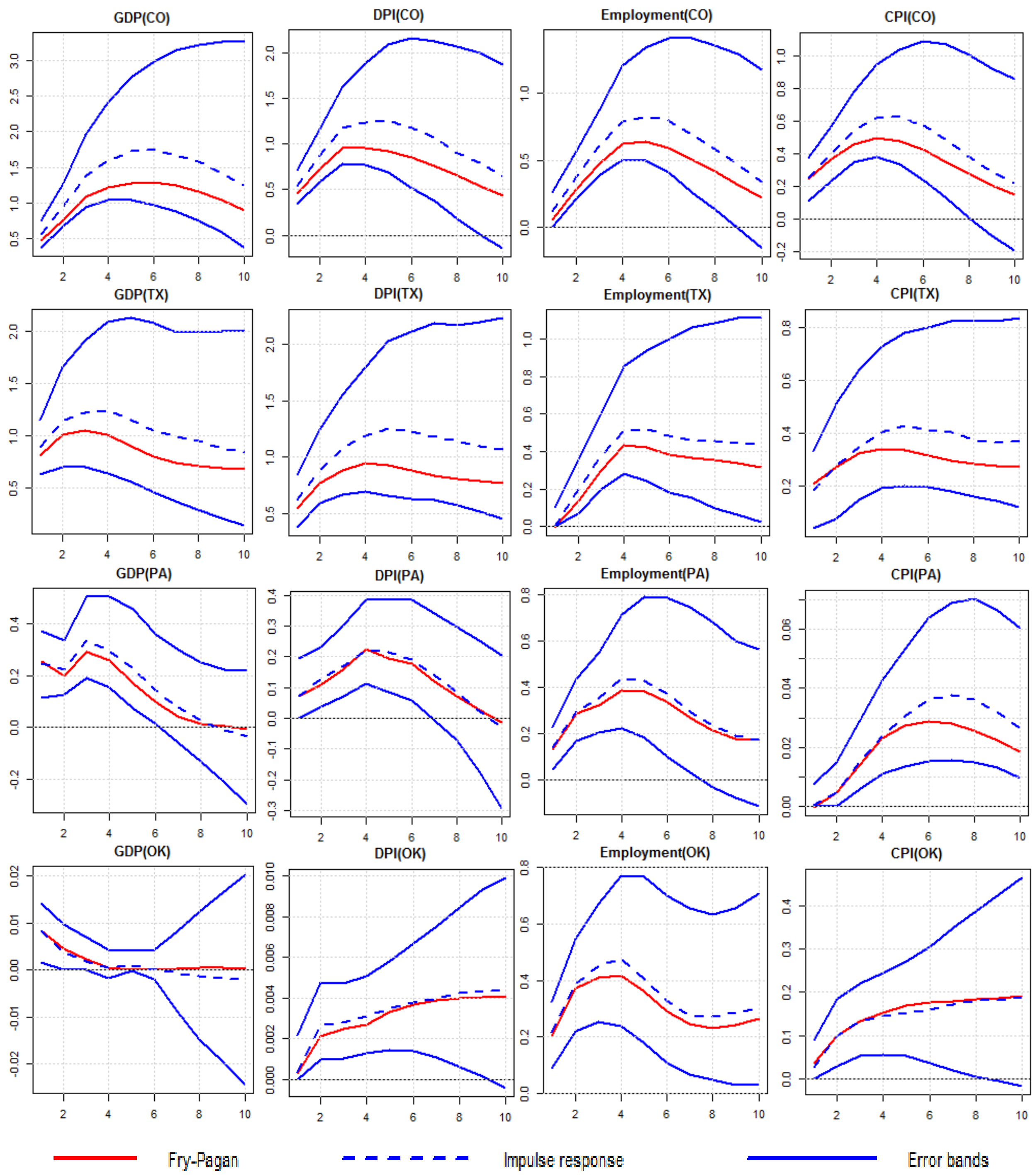

Figure 7: Comparison of impulses responses (GDP). The 1st (CO), 2nd (TX), 25th (PA), and 50th (OK) state's GDP with state's disposable personal income, CPI, and employment responses are drawn from Fry and Pagan's (2011) M-T method (solid red line) and sign restricted FAVAR models (dashed blue line). The GDP responses of 1st (CO), 2nd (TX), 25th (PA), and 50th (OK) states are ordered by their cumulative impulse responses over a 10-year horizon. Solid blue lines represent the 95 % confidence intervals. Tax shock is measured as a 1 percent cut in the federal personal income tax rate.

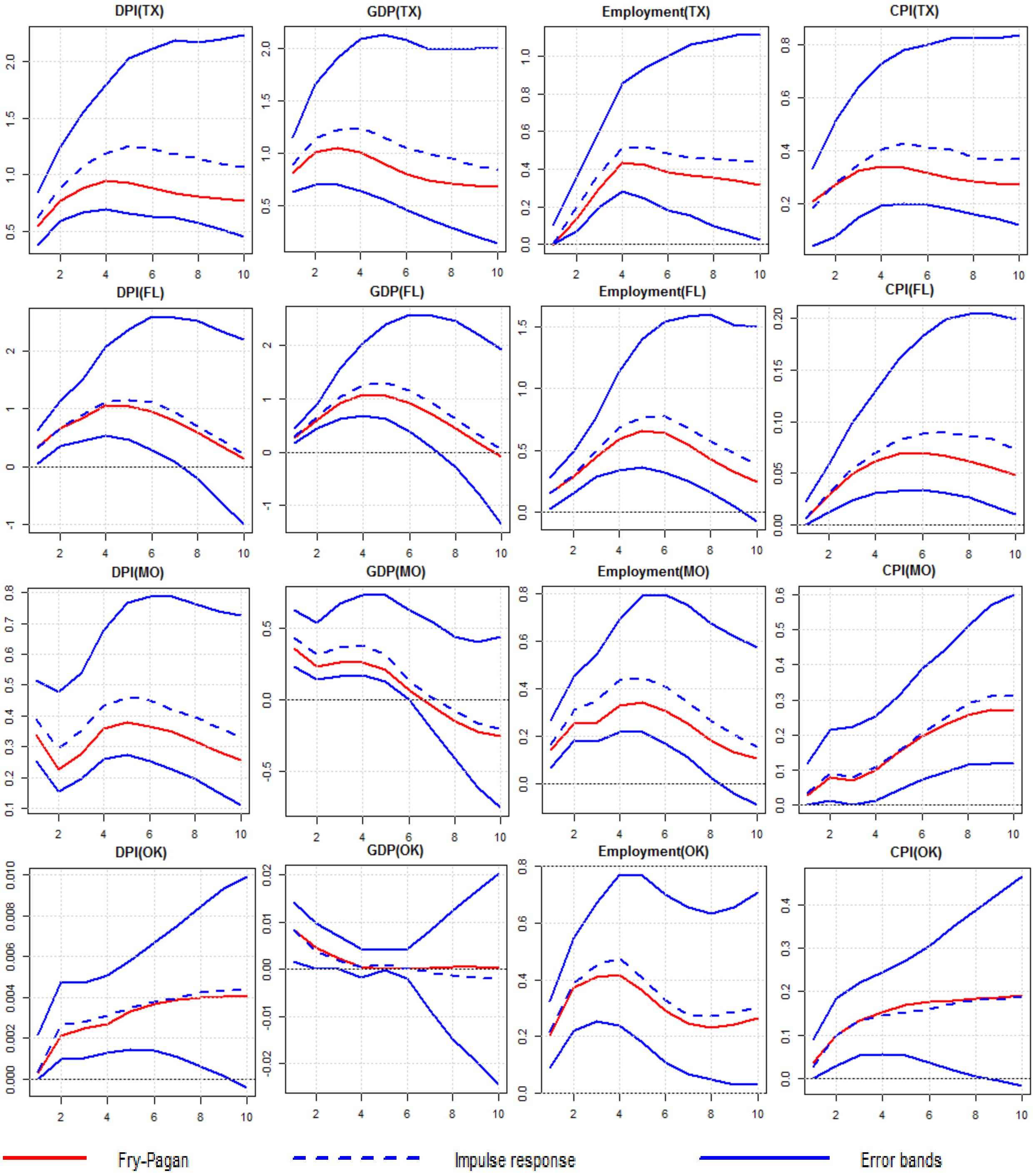

Figure 8: Comparison of impulses responses (DPI). The 1st (TX), 2nd (FL), 25th (MO), and 50th (OK) state's DPI with those state's GDP, CPI and employment responses are drawn from Fry and Pagan's (2011) M-T method (solid red line) and sign restricted FAVAR models (dashed blue line). The DPI responses of $1^{st}$ (TX), $2^{nd}$ (FL), $25^{th}$ (MO), and $50^{th}$ (OK) states are ordered by their cumulative impulse responses over a 10-year horizon. Solid blue lines represent the 95 % confidence intervals. Tax shock is measured as a 1 percent cut in the federal personal income tax rate.

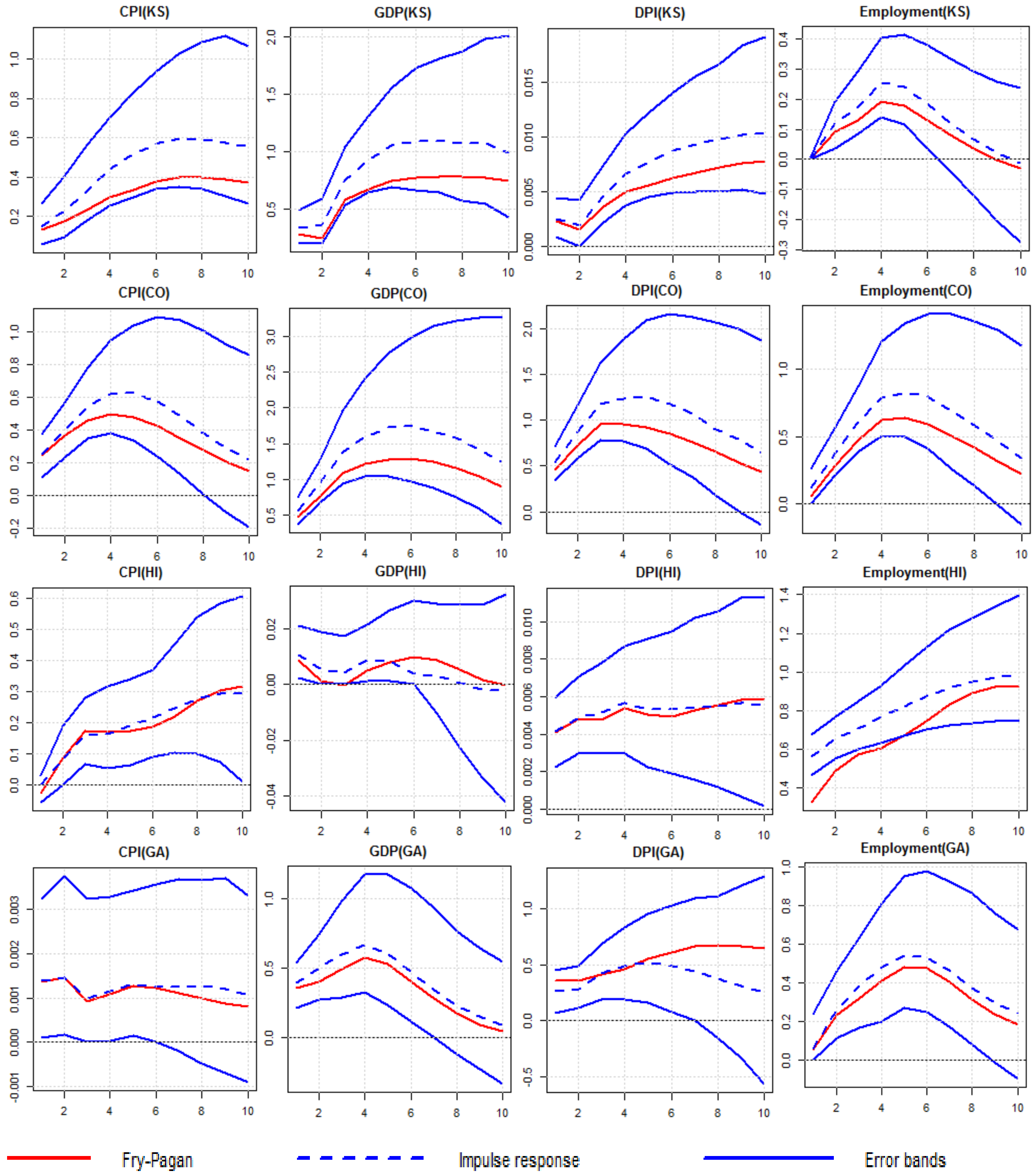

Figure 9: Comparison of impulses responses (CPI). The 1st (KS), 2nd (CO), 25th (HI), and 50th (GA) state's CPI with those state's GDP, DPI and employment responses are drawn from Fry and Pagan's (2011) M-T method (solid red line) and sign restricted FAVAR models (dashed blue line). The CPI responses of 1st (KS), 2nd (CO), 25th (HI), and 50th (GA) states are ordered by their cumulative impulse responses over a 10-year horizon. Solid blue lines represent the 95 % confidence intervals. Tax shock is measured as a 1 percent cut in the federal personal income tax rate.

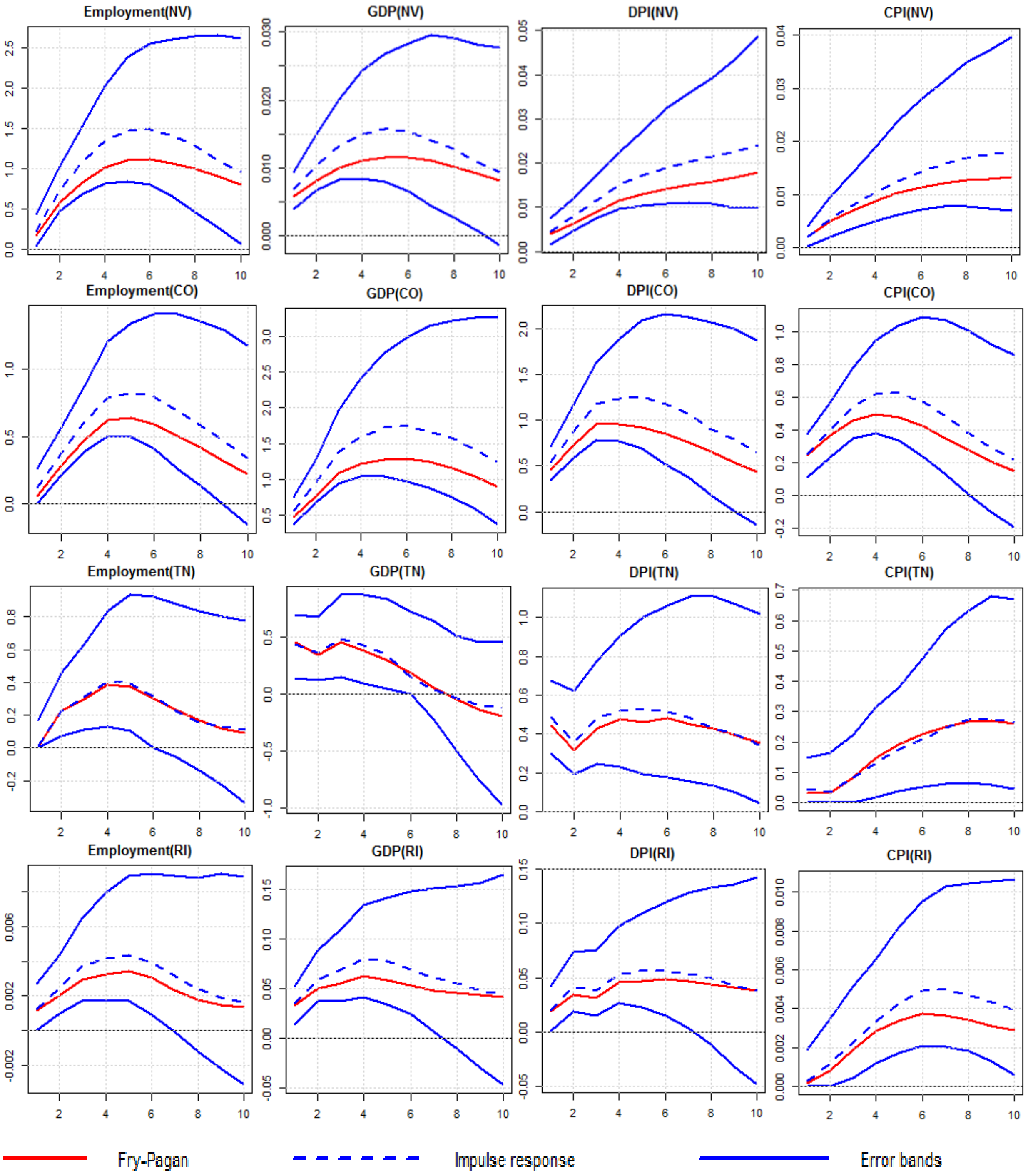

Figure 10: Comparison of impulses responses (employment). The 1st (NV), 2nd (CO), 25th (TN), and 50th (RI) state's employment with those state's GDP, DPI, and CPI responses are drawn from Fry and Pagan's (2011) M-T method (solid red line) and sign restricted FAVAR models (dashed blue line). The employment responses of 1$^{st}$ (NV), 2$^{nd}$ (CO), 25$^{th}$ (TN), and 50$^{th}$ (RI) states are ordered by their cumulative impulse responses over a 10-year horizon. Solid blue lines represent the 95 % confidence intervals. Tax shock is measured as a 1 percent cut in the federal personal income tax rate.

### 3.6 Economic significance of the FAVAR estimates

In general, the findings of this study support the short-term prediction of the expansionary effect of tax cuts on real GDP, employment, and prices through the aggregate demand (AD) channel of the New Keynesian model. The cut in taxes increases real GDP in this model in the short run because corporate tax cuts encourage firms to increase investment, and personal income tax cuts increase disposable incomes, increasing demand for goods and services. Specifically, on impact, a one percent cut in personal income tax increases real GDP for 33 states by an average of 1.2 percent with maximum of 1.9 percent in five years. Similarly, a one percent cut in corporate income tax cut raises real GDP for 40 states by 0.52 percent on impact and by 0.83 percent in three years. The economic interpretation of these results is related to the national income multiplier. The income multiplier suggests how big a percentage change in real GDP is due to a one percent change in both taxes in the FAVAR model.

Interpreted as a multiplier in the Keynesian model, the FAVAR estimates imply a maximum personal income tax's output multiplier of 1.9, and the corporate income tax's output multiplier is 0.83. These findings support several previous studies related to the variation in estimates of the U.S. fiscal multipliers. For example, Barro and Redlick (2011) estimate a U.S. personal income tax multiplier of around 1.1, Mountford and Uhlig (2009) find 0.93 after one year of the tax cut, and up to 3.41 at three years, Riera-Crichton et al. (2016) find a tax multiplier of 1.32 using the SVAR approach and 2.76 using the narrative approach for 14 industrial countries. The estimates from the Congressional Budget Office show output multipliers for the lower- and middle-income group is in between 0.3 and 1.5 (Whalen and Reichling, 2015).

At the state level, the value of output multiplier starts rising two years after the personal income tax cut for most states and reaches a maximum of 1.6 for Colorado, 1.2 for Texas, and 0.38 for Pennsylvania at five years. The multipliers decrease gradually over the next five-year horizons. The value of multipliers is much lower at longer horizons because of the "crowding out" of private investment. In the long run, tax cuts increase the federal budget deficit, and the government's increased borrowing "crowds out" state-level private investment.

A one percent change in personal income tax raises disposable personal income by a maximum of 1.2 percent and a minimum of 0.05 percent. A common economic interpretation for the maximum effects is the peak personal income multiplier.[3] Since the impact response of $PIT_t$ to the tax cut shock is rescaled to unity, the economic significance of the impulse responses of disposable personal income at a 10-year horizon is the 10-year multiplier. The total accumulated responses for 10-year periods ahead of a unit tax cut shock on $DPI_t$ can be calculated by summing up the corresponding response coefficients. This accumulated value is commonly referred to as the total multiplier effect. As expected, the total effect is higher for CO, TX, FL, and MD with an average of 1.115, followed by an average of 0.79 for KS, AZ, IA, MN, MA, WA, and CA.

Looking at the labor market, a one percent cut in personal income tax increases employment by a maximum of 1.6 percent compared to 0.92 percent for the corporate income tax cut. Translated into the non-farm employment elasticity,[4] the FAVAR estimates of elasticities are economically consistent with the previous studies (Mourre, 2006; Kapsos, 2006; Seyfried, 2011). Table 5 reports state-level employment elasticities. The values reveal that the higher non-farm employment response growth is observed in CO, NV, AZ, FL, ID, TX, CA, WA, NJ, and GA. Examining the state's higher employment growth and their relatively higher output multiplier also reveals that the tax cut's expansionary effects have worked simultaneously in the short-run for most states. The employment elasticity continued to rise from 0.001 to 1.6 until around five years after the cut in either taxes. The average employment elasticity of this study is about 0.424 for the personal income tax changes and 0.295 for the corporate tax changes. These estimates are close to Mourre (2006), who estimates the economy-wide employment elasticity is 0.6. Seyfried (2011) also estimates employment elasticity range from 0.31 to 0.61for the ten specific states, with the U.S. average estimate of 0.47.

[3] The multiplier is $sign(d \ln DPI_{t+h} / d \ln PIT_t)$, where $DPI_t$ is disposable personal income at time $t$, $PIT_t$ is personal income tax, $h$ is the 10-years of forecasts horizon.

[4] Employment elasticity measures the ratio of the percentage change in the number of non-farm employment in a state to the percentage change in narrative tax rates: $\varepsilon_{emp} = (\%\Delta\, in \quad nonfarm \quad emp / \%\Delta\, in \quad tax)$

Table 5: Nonfarm employment elasticity

| State | PIT | CIT | State | PIT | CIT |
|---|---|---|---|---|---|
| AL | 0.31 | 0.18 | MT | 0.005 | 0.004 |
| AK | 0.015 | 0.002 | NC | 0.4 | 0.18 |
| AR | 0.21 | 0.23 | ND | 0.012 | 0.01 |
| AZ | 0.81 | 0.57 | NE | 0.23 | 0.16 |
| CA | 0.61 | 0.37 | NH | 0.005 | 0.003 |
| CO | 1.2 | 1.28 | NJ | 0.45 | 0.18 |
| CT | 0.41 | 0.24 | NM | 0.42 | 0.28 |
| DE | 0.5 | 0.36 | NV | 1.5 | 0.92 |
| FL | 0.8 | 0.48 | NY | 0.41 | 0.12 |
| GA | 0.6 | 0.21 | OH | 0.43 | 0.38 |
| HI | 0.6 | 0.012 | OK | 0.45 | 0.18 |
| IA | 0.53 | 0.51 | OR | 0.42 | 0.33 |
| ID | 0.91 | 0.43 | PA | 0.41 | 0.25 |
| IL | 0.38 | 0.24 | RI | 0.005 | 0.003 |
| IN | 0.31 | 0.22 | SC | 0.42 | 0.21 |
| KS | 0.34 | 0.43 | SD | 0.43 | 0.41 |
| KY | 0.42 | 0.4 | TN | 0.41 | 0.26 |
| LA | 0.005 | 0.004 | TX | 0.51 | 0.23 |
| MA | 0.4 | 0.72 | UT | 0.61 | 0.45 |
| MD | 0.4 | 0.22 | VA | 0.58 | 0.18 |
| ME | 0.42 | 0.32 | VT | 0.41 | 0.3 |
| MI | 0.5 | 0.48 | WA | 0.51 | 0.4 |
| MN | 0.21 | 0.32 | WI | 0.41 | 0.32 |
| MO | 0.44 | 0.37 | WV | 0.001 | 0.002 |
| MS | 0.42 | 0.4 | WY | 0.015 | 0.001 |

## 4. Understanding state heterogeneity in the transmission of tax shocks

As evidenced in section 3.4, state-level responses vary substantially over time and across state. This variation raises the question of what state-level structural characteristics drive the heterogeneous responses of a tax cut shock? To begin to address this I regress the state-level FAVAR results to the state sectoral contribution to their GDP, fiscal and labor market structures, financial markets, and the housing sector.

### 4.1 Cross-state regression

With the state-level cumulative impulse responses and the cross-state characteristics, the regression model (2) estimates state-level characteristics that drive the heterogeneous responses:

$$Cum\,\mathrm{Re}_{t,m}^{t} = \varphi + \beta X_i + \varepsilon_i \tag{2}$$

where $Cum\,\mathrm{Re}_{t,m}^{t}$ indicates the $t-year$ horizon cumulated impulse responses of state $i's$ macroeconomic variable $m \in [real\,GDP, personal\,income, CPI, employment]$. In contrast to the impact of tax cut shock at a single point in time, the cumulative impulse responses ($Cum\,\mathrm{Re}_{t,m}^{t}$) examine the total accumulated effect of the impact of the tax cut shock to one of the four macroeconomic variables in *m* across ten-year horizons. The cumulative impulse responses have been broadly applied to examine, among others, government spending multipliers (Owyang, Ramey, and Zubairy, 2013), the macroeconomic impact of the oil price shock (Hamilton, 2011), and the output effects of technology shocks (Gali, 1999). Apart from these applications, the cumulative value of impulse response functions is also used by Carlino and DeFina (1998, 1999), Liu and Williams (2019), and Owyang and Zubairy (2013) as a dependent variable in the cross-section regression model.

The regressors $X_i$ include variables that attempt to explain the cross-state heterogeneities in the financial sector, industry share in state's GDP, fiscal conditions, and the labor and housing market. Motivated by previous literature, the first set of covariates comprises the sectoral composition of the state's GDP and the median household income. From 1977 to 2018, the state's sectoral composition measures as the percentage of the state's GDP accounted for by agriculture, financial, professional, business services, manufacturing, oil and mining, housing, construction, and the public sector.

#### 4.1.1 Sectoral composition

One possible reason states display heterogeneity in their responsiveness is the diversity of the sectoral composition of the state's aggregate output. For example, states with a higher GDP concentration accounted for by durable and non-durable consumer goods might show a higher response to a personal income tax cut. As in the standard Keynesian model, states output, employment, and price levels should also display a strong positive response to this expansionary tax cut shock through the aggregate demand and income channel. Similarly, states with a larger concentration of capital-intensive industries or production of durable goods might be more responsive to a corporate tax cut. Barth and Ramey (2001) show that corporate tax cut shocks transmit to the state's industrial and manufacturing sectors through the cost channel of firm finance and encourage firms to take advantage of debt finance options rather than equity (Heckemeyer and Mooij, 2017). Several empirical studies, particularly in the response of monetary and fiscal policy shocks, find that the composition of GDP share accounted for by state's manufacturing and financial sector can explain the heterogeneous responses and the transmission mechanisms (Bernanke and Gertler, 1995; Carlino and Defina, 1998; Owyang and Zubairy, 2013).

Figure 11 shows the concentration of private industries, financial, and manufacturing share as a percentage of the state's GDP. As shown in the bar-graph, there has been significant variation in the states' financial and manufacturing sectors. For example, South Dakota, Iowa, and Delaware have the highest concentration of the financial industry in their GDP, while Idaho and Oregon are noticeably more dependent on the manufacturing sector. The states located in the South, such as Georgia, North Carolina, South Carolina, have a small share of manufacturing and financial industries in their GDP.

#### 4.1.2 Financial sector

As shown in the bottom panel of Figure 11, states differ significantly in the concentration of financial industries. These differences can shed light on the analysis of the cross-states heterogeneities of tax shocks transmission. Mooij and Hebous (2018) show that the magnitude of debt finance elasticity to a corporate tax cut significantly correlated with a firm's asset size, where relatively smaller and larger companies show higher response to tax changes. So, corporate tax cuts can have higher and quicker responses in states where they are located. The recent empirical evidence, as in Fernández-Villaverde (2010), Gertler and Kiyotaki (2010), and Zwick and Mahon (2014) also emphasize the importance of financial frictions in examining the effects of fiscal policy changes.

Following Carlino and Defina (1998) and Cottarelli and Kourelis (1994), the second set of covariates include total bank loans to households and businesses, loans to state-level small firms, asset and equity for commercial banks, net interest margins, and net income for commercial banks. The proportion of bank loans to households and businesses is the proxy measure of broad credit channel, which assumed to be strongly correlated to the response of output, income, and prices (Dornbusch et al., 1998; Mihov, 2001). The state-level bank's asset, equity, and net income for commercial banks are the proxy measure of the financial sector's response to tax changes. Finally, the bank's net interest margin is a proxy of the banking sector's efficiency.

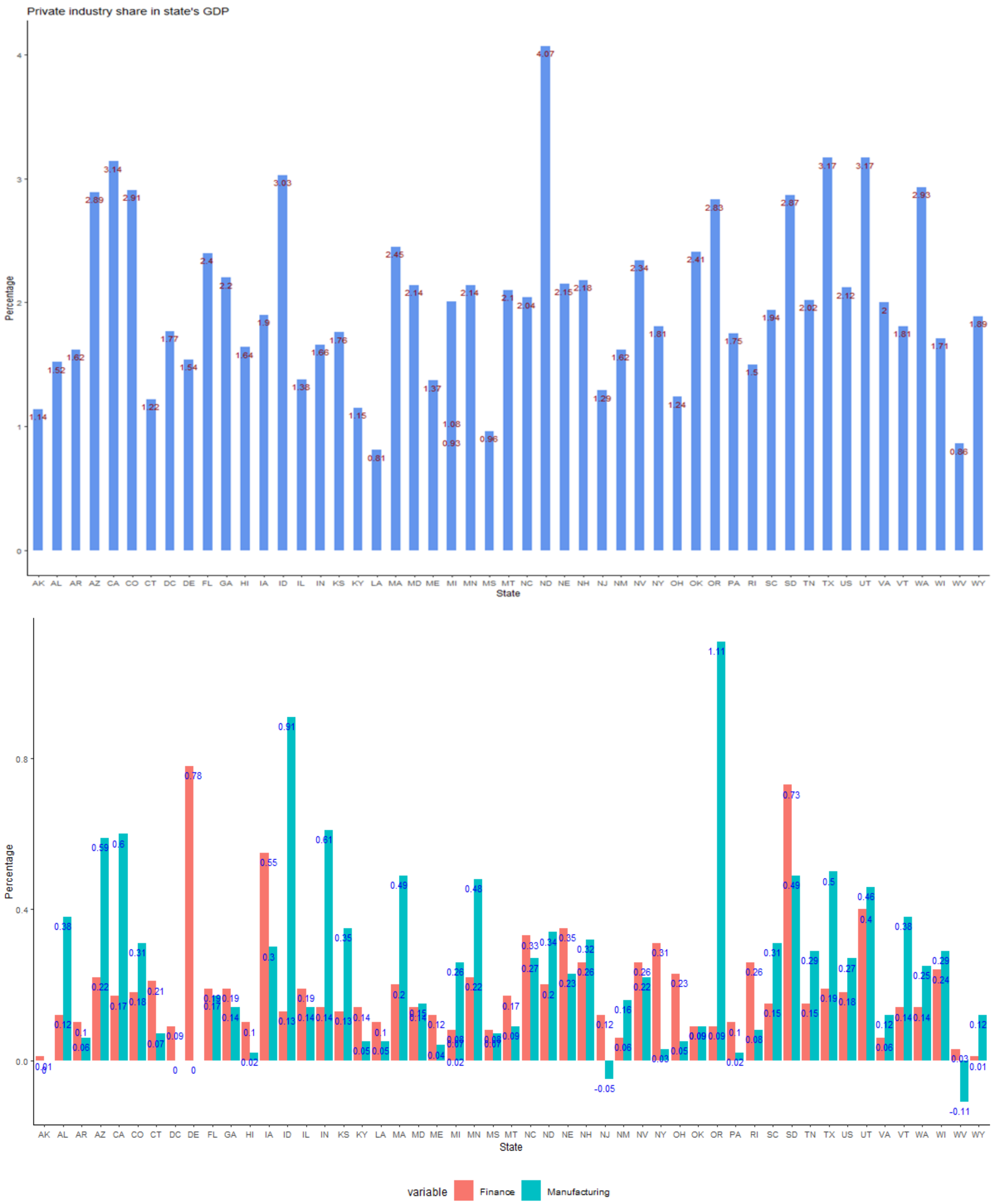

Figure 11: State Heterogeneity of sectoral composition. Top panel shows share of private industries in state GDP (1977-2018); the bottom panel shows the share of manufacturing and financial sector in state's GSP (1977-2018).

#### 4.1.3 Fiscal structure

The fiscal structures of U.S. states vary substantially across the states, as shown in Figure 12. The top panel of Figure 12 presents the state's average values and local tax burdens, and the bottom panel shows the state's income tax elasticity[5]. California, Missouri, Illinois, New York, Minnesota, and North Carolina are characterized as states with the highest tax burden, while Wyoming, Nevada, South Dakota, and Texas have the lowest tax burdens. Following Liu and Williams (2019) and Owyang and Zubairy (2013), this study's state-level fiscal policy variables are divided into two major categories. The first category consists of state-specific tax variables, such as state and local tax burdens, the elasticity of personal income tax, average personal income, and corporate income tax rate.  The second group includes state tax revenue, real sector uncertainty, subsidy, and state unemployment benefits. Many of these series are taken from Mumtaz et al. (2018) and the annual survey of state and local government finances of the U.S. Census Bureau.

#### 4.1.4 Housing and Labor markets

The consensus in the empirical literature is that the more rigid labor markets strongly associated with the more robust and persistent responses of output, prices, and employment to a policy change (Brückner and Pappa, 2012; Walsh, 2005; Zanetti, 2007).  For example, Zanetti (2007) finds that union bargaining reduces the elasticity of marginal costs to output and increases output. This study includes state-level unemployment rate, civilian labor force participation, and sectoral employment to document the importance of labor market frictions. Besides, the proxy of “Right to Work” laws and the degree of unionization (Mumtaz et al., 2018) measure the labor market rigidities. The “Right to work” laws exist in 28 states. The labor market structure in those states is flexible because the law allows workers to make their own choice if they want to join a union. The regressors' final set is from the state’s housing sector, such as homeownership rate, home vacancy rate, rental vacancy rate, and percentage of the housing sector to the state’s GDP.

[5] Income tax elasticity measures the ratio of the percentage change in tax liabilities to a percentage change in state’s income ( Snowbarger and Kirk, 1973)

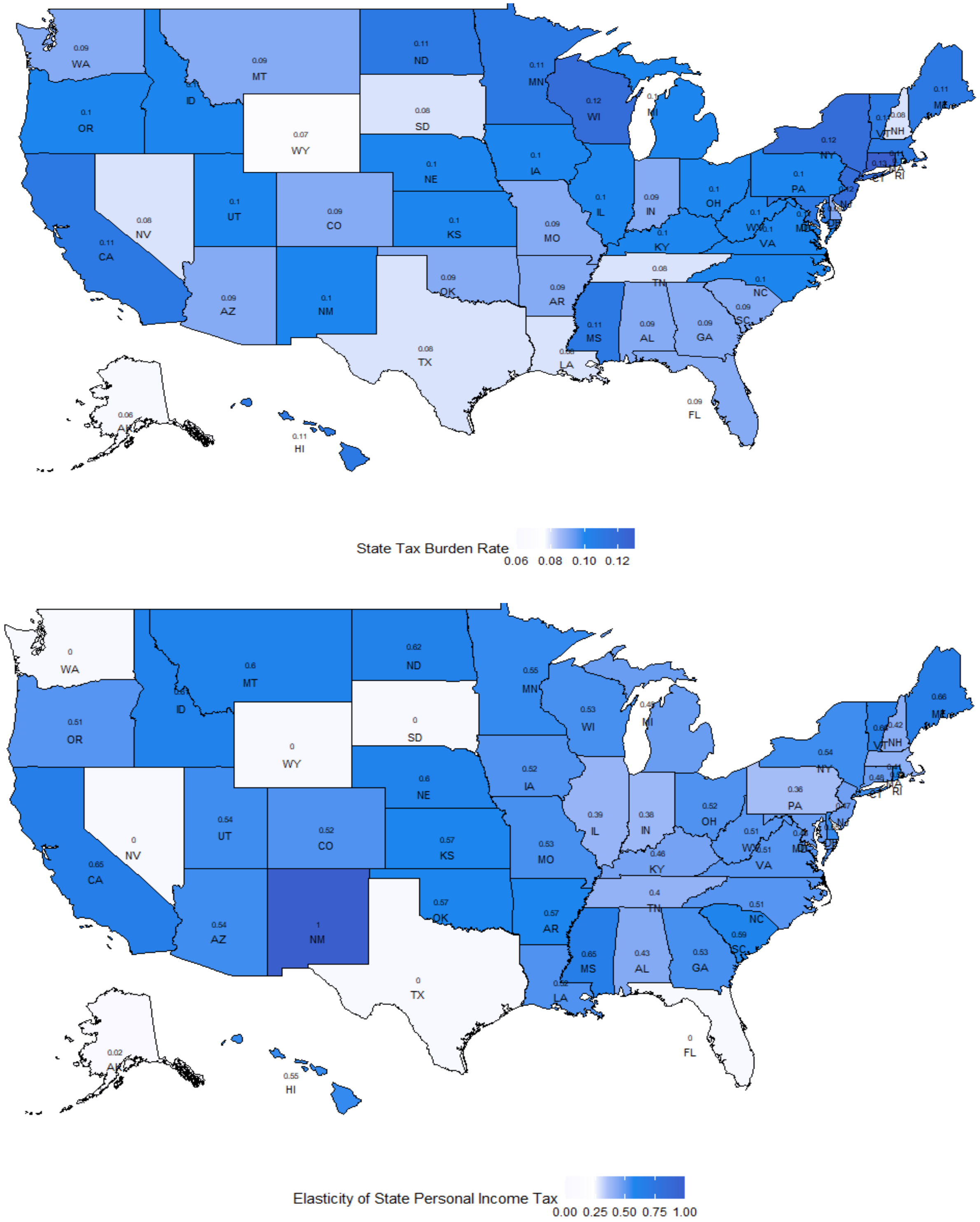

Figure 12: Distribution of state tax burden and tax elasticity. Top panel shows the statewide tax burden rate; the bottom panel shows the statewide elasticity of personal income tax (average for 1984-2018). Values are normalized.

### 4.2 Cross-state regression results: GDP

Table 6 shows the results of the regression model (2) estimated by OLS. This model includes the key regressors explained in sections 4.1.2 through 4.1.4. The specification of the model and the inclusion of regressors broadly based on previous empirical literature (as in Carlino and Defina 1998; Oyang and Zubuary, 2013; Liu and Williams 2019). The first column shows the results of the regressions for the impulse response of GDP cumulated at the 10-year horizons, while GDP responses cumulated at the 2-year horizons is in second column. The dependent variable in column one refers to long-run responses, and the Cumulative-IRF 2yr indicates the short-run responses. The model includes two fiscal variables, manufacturing and financial sector share, state-level tax elasticity, variable from labor markets, housing sector, and financial markets.

The cross-state regression analysis applies the standardized regression approach that standardized state-specific all covariates on the scale of zero mean and one standard deviation. The motivation of the standardized regression approach is to determine sectoral relative strength to understand heterogeneous impulse responses. So, the interpretation of the regression results in Table 6 is the standardized regression coefficients that are based on changes in standard deviation units. For example, for every standard deviation unit increase in the share of financial and manufacturing in the state's GDP in column (1) of Table 6, the states' real GDP response increases by 0.59 and 0.38 standard deviation units, respectively. Comparing the relative importance of these two sectors, these values show that a larger ratio of financial industries in the state's GDP is related to a higher positive response of real output than the manufacturing industries. Similarly, the negative coefficient in Table 6 shows that the state's dependency on the debt has the most adverse effect on the output response (-0.34) followed by state policy uncertainty (-0.27) and strict labor market regulation (-0.24).

As shown in column (1) and (2), the state-level covariates display a robust relationship between the cumulative responses of personal income tax shocks and the state-level economic structures. For the long-run responses, column (1) shows that the financial and manufacturing

Table 6: Cross-state regression results: personal income tax

| | GDP | |
|---|---|---|
| | (1) Cumulative-IRF10 yr | (2) Cumulative-IRF 2 yr |
| Financial | 0.59*** | 0.65*** |
| | (0.16) | (0.16) |
| State Govt. debt | -0.34** | -0.39** |
| | (0.15) | (0.15) |
| State uncertainty | -0.27** | -0.19 |
| | (0.13) | (0.13) |
| Manufacturing | 0.38** | 0.49*** |
| | (0.14) | (0.14) |
| State job creation | 0.32** | 0.32** |
| | (0.12) | (0.12) |
| Loans to small firms | 0.26* | 0.21* |
| | (0.13) | (0.13) |
| Elasticity of income tax | -0.31** | -0.30** |
| | (0.14) | (0.14) |
| Home ownership rate | -0.36** | -0.35** |
| | (0.15) | (0.15) |
| State Avg Corp.Inc. Tax | 0.001 | -0.014 |
| | (0.15) | (0.15) |
| Labor market regulation | -0.24* | -0.24* |
| | (0.13) | (0.13) |
| Observations | 50 | 50 |
| Adjusted $R^2$ | 0.39 | 0.39 |

Note: *p<0.1; **p<0.05; ***p<0.01

share have higher statistically significant positive effect on the expansionary response of state-level GDP. The positive coefficient on the share of finance suggests that the higher concentration of finance, insurance, and banks increase the GDP in the face of federal tax cuts.

Similarly, states with a higher share of manufacturing industries also experience a positive GDP response. The growth of state-level non-farm job creation enters with a statistically significant positive coefficient in both the short- and long-run, reflecting the persistence response in the state's GDP growth. The homeownership rate has a negative coefficient, as explained in section 5.4, significant at the 5 percent level, the percentage of bank loans to small firms has a positive effect on the response of GDP, and negatively related with the increase in state income tax rates. The state corporate income tax also shows an adverse effect, but coefficients are statistically insignificant in both the short- and long-run.

The positive correlation between state financial and manufacturing share and the real GDP response to a corporate tax cut is also positive and statistically significant across two regressions, as shown in Table 7. As in the response of personal income tax cuts, states with higher job creation rates tend to have higher real GDP responses to the corporate tax cuts. The responses are negatively correlated to the state homeownership rate, while state and local government debt are also negative but statistically insignificant.

The findings in Table 6 and 7 have a connection to the impulse responses in section 3.4. The personal income tax cut leads to a higher increase in real GDP in states, like Texas, Colorado, Oregon, Idaho, California, and Florida. As shown in figure 11, the concentration of private industries, particularly finance and manufacturing share, is also higher in these six states. For example, private industry share accounts for a maximum of 3.87 percent in Texas, followed by 3.14 percent in California and 3.03 percent in Idaho. So, the larger share of financial and manufacturing industries for these states supports the regression results in Tables 6 and 7.

The effect of labor market regulation on the response of GDP is negative and statistically significant. The magnitude of the estimated coefficients is the same for both short- and long-run horizons, and the effects are robust. The coefficient of state-level real sector uncertainty is also negative and significant, suggesting state with higher uncertainty may signal an uncertain future

Table 7: Cross-state regression results: corporate income tax

| | GDP | |
|---|---|---|
| | (1) | (2) |
| | Cumulative-IRF10 yr | Cumulative-IRF 2 yr |
| Financial | 0.53** | 0.42*** |
| | (0.23) | (0.27) |
| Manufacturing | 0.41** | 0.48*** |
| | (0.16) | (0.16) |
| State uncertainty | -0.22 | -0.13 |
| | (0.15) | (0.14) |
| State Avg Corp.Inc. Tax | -0.10 | -0.09 |
| | (0.17) | (0.17) |
| Elasticity of income tax | -0.25* | -0.28* |
| | (0.16) | (0.16) |
| Loans to small firms | 0.01 | 0.005 |
| | (0.15) | (0.14) |
| State job creation | 0.26* | 0.20 |
| | (0.15) | (0.14) |
| Home ownership rate | -0.40** | -0.32* |
| | (0.17) | (0.17) |
| Labor market regulation | -0.02 | -0.04 |
| | (0.15) | (0.15) |
| Housing | -0.30 | -0.18 |
| | (0.22) | (0.21) |
| State Govt. debt | -0.11 | 0.18 |
| | (0.17) | (0.17) |
| Observations | 50 | 50 |
| Adjusted $R^2$ | 0.20 | 0.22 |

Note: *p<0.1; **p<0.05; ***p<0.01

environment when households and businesses are unwilling to incorporate the tax cut incentives in their utility and production functions (Bloom, 2009; Mumtaz et al., 2018; Bloom et al., 2018). However, the coefficients of state uncertainty index, fiscal structure, and labor market regulation do not appear as a significant and robust component in explaining the heterogeneity of GDP responses to corporate income tax changes.

Additionally, regression models include state government debt as a percentage of Gross State Product (GSP), which is considered a key measure of state and local government fiscal structure. The coefficient is negative and statistically significant across all models. The rising debt to GSP ratio indicates a state's weaker fiscal conditions and plays a negative role on the long-run response of GDP to a tax cut shock. This finding partially supports the critique of the "over-borrowing" model, which primarily depends on the principle of the Ricardian Equivalence Model. The model assumes that higher government debt and future debt payments translate into lower consumption, more private savings, and lower property values (Barro, 1974; Eichenberger and Stadelmann, 2010). As a result, real output and prices decline as in the New Classical and New Keynesian Model through aggregate demand channel components. This view supports the findings of Banzhaf and Oats (2012), who study the Ricardian Equivalence Model for U.S. states. They find that local debt finance options increase future tax burden and translate into lower property values and income.

### 4.3 Cross-state regression results: DPI, employment, and CPI

Column (1) through (4) in Table 8 and 9 show the estimated results of the regression models for the cumulative response of personal income and non-farm employment. The response of personal income and employment at 10- and 2-year horizons are the dependent variable in both tables. The financial and manufacturing share, job creation rate, and loan to small firms show a robust and positive effect on personal income and non-farm employment. State labor market regulation, government debt, and state-level tax rates are negatively related to the response of employment. The economic policy uncertainty also shows a negative impact on the response of these two variables, but coefficients are statistically insignificant.

Table 8: Cross-state regression results: personal income tax

| | Dependent variable | | | |
|---|---|---|---|---|
| | (1) | (2) | (3) | (4) |
| | IRF(DPI)10 yr | IRF(DPI) 2 yr | IRF (EMP) 10 yr | IRF(EMP) 2 yr |
| Financial | 0.59*** | 0.62*** | 0.76*** | 0.78*** |
| | (0.15) | (0.16) | (0.17) | (0.16) |
| State Govt. debt | -0.37** | -0.37** | -0.33** | -0.34** |
| | (0.15) | (0.15) | (0.15) | (0.13) |
| Manufacturing | 0.44*** | 0.46*** | 0.29* | 0.31* |
| | (0.14) | (0.14) | (0.17) | (0.17) |
| State job creation | 0.17 | 0.23* | 0.33** | 0.27* |
| | (0.12) | (0.12) | (0.13) | (0.14) |
| Loans to small firms | 0.37*** | 0.27** | -0.006 | -0.03 |
| | (0.13) | (0.13) | (0.13) | (0.13) |
| Elasticity of income tax | -0.34** | -0.34** | -0.27 | -0.24 |
| | (0.14) | (0.15) | (0.18) | (0.18) |
| Home ownership rate | -0.37** | -0.34** | -0.10 | -0.004 |
| | (0.15) | (0.15) | (0.16) | (0.15) |
| Labor market regulation | -0.36*** | -0.38*** | -0.34** | -0.29** |
| | (0.13) | (0.13) | (0.14) | (0.13) |
| State uncertainty | -0.14 | -0.17 | -0.17 | -0.14 |
| | (0.13) | (0.13) | (0.13) | (0.13) |
| State Avg Corp. Inc. Tax | 0.05 | 0.01 | 0.13 | 0.21 |
| | (0.15) | (0.15) | (0.18) | (0.19) |
| Observations | 50 | 50 | 50 | 50 |
| Adjusted $R^2$ | 0.39 | 0.38 | 0.329 | 0.35 |

Note: *p<0.1; **p<0.05; ***p<0.01

Table 9: Cross-state regression results: corporate income tax

| | Dependent variable | | | |
|---|---|---|---|---|
| | (1) IRF (DPI) 10 yr | (2) IRF (DPI) 2 yr | (3) IRF(EMP) 10 yr | (4) IRF(EMP) 2 yr |
| Financial | 0.48** | 0.51*** | 0.34* | 0.39** |
| | (0.17) | (0.17) | (0.20) | (0.19) |
| State Govt. debt | -0.34** | -0.36** | -0.04 | -0.15 |
| | (0.16) | (0.16) | (0.19) | (0.17) |
| Manufacturing | 0.48*** | 0.42** | 0.18* | 0.30* |
| | (0.16) | (0.16) | (0.18) | (0.17) |
| State job creation | 0.22 | 0.27* | 0.02 | 0.04 |
| | (0.14) | (0.14) | (0.16) | (0.15) |
| Labor market regulation | -0.33** | -0.32** | -0.23 | -0.03 |
| | (0.15) | (0.14) | (0.17) | (0.15) |
| Loans to small firms | 0.016 | -0.04 | 0.22 | 0.08 |
| | (0.14) | (0.14) | (0.16) | (0.15) |
| Elasticity of income tax | -0.40** | -0.40** | 0.09 | 0.14 |
| | (0.16) | (0.16) | (0.18) | (0.18) |
| Home ownership rate | -0.22 | -0.17 | -0.08 | 0.13 |
| | (0.17) | (0.17) | (0.16) | (0.18) |
| State uncertainty | -0.11 | -0.14 | -0.17 | -0.14 |
| | (0.14) | (0.15) | (0.13) | (0.13) |
| State Avg Corp. Inc. Tax | 0.11 | 0.12 | -0.14 | -0.14 |
| | (0.17) | (0.17) | (0.20) | (0.19) |
| Observations | 50 | 50 | 50 | 50 |
| Adjusted $R^2$ | 0.23 | 0.25 | 0.38 | 0.34 |

Note: *p<0.1; **p<0.05; ***p<0.01

Regression models in Table 10 augment the regression models by adding four financial friction-related variables: loans to households and business, the ratio of loan loss to the bank's total asset, the ratio of non-performing loans to total bank's loan, and net interest margin. The motivation of the inclusion of these additional variables comes from the empirical framework of Carlino and Defina (1998, 1999), Di Giacinto (2003), Cecchetti (2001), and Lucio and Izquierdo (1999). The objective is to examine the influence of state-level liquidity and money supply variables on the response of price levels. Corporate tax cuts increase banks and other lenders' liquid assets, and thus financial institutions respond to tax cuts by providing more loans to consumers and firms (Bernanke 1981; Buiter and Panigirtzoglou, 1999).

The regression results in Table 10 support this liquidity effect, where loans to small firms and loans to households and businesses show a robust and positive effect on the response of consumer price index. The ratio of loan loss to total loan, non-performing loans, and the net interest margins appears as contractionary components. In general, these components reduce bank solvency and signals increased uncertainty about future liquidity. The coefficient of non-performing loans is negative, suggesting adverse effect on the response of in price levels. The ratio of loan loss to total assets is not robust across the models.

Table 10: Cross-state regression results: CPI

| | Dependent variable | | | |
|---|---|---|---|---|
| | (PIT Shock) | | (CIT Shock) | |
| | (1) | (2) | (3) | (4) |
| | IRF (CPI) 10 yr | IRF(CPI) 2 yr | IRF(CPI) 10 yr | IRF(CPI) 2 yr |
| Financial | 0.38** | 0.29* | 0.39** | 0.37** |
| | (0.16) | (0.17) | (0.16) | (0.16) |
| Manufacturing | 0.28 | 0.30* | 0.26 | 0.29* |
| | (0.17) | (0.17) | (0.16) | (0.16) |
| Loans to small firms | 0.49*** | 0.38** | 0.42** | 0.42** |
| | (0.17) | (0.17) | (0.17) | (0.17) |
| Loans to HH & Business | 0.98** | 1.12** | 1.17*** | 1.31*** |
| | (0.46) | (0.47) | (0.44) | (0.45) |
| Loan loss/total asset | -0.22 | -0.17 | -0.22 | -0.34 |
| | (0.21) | (0.21) | (0.21) | (0.21) |
| Non-perform loans/total loans | -0.68 | -0.17 | -0.83* | -0.94* |
| | (0.44) | (0.45) | (0.43) | (0.43) |
| Net int. margin | -0.20 | -0.25 | -0.29 | -0.24 |
| | (0.18) | (0.18) | (0.17) | (0.17) |
| State job creation | 0.09 | 0.12 | 0.01 | 0.09 |
| | (0.15) | (0.15) | (0.14) | (0.14) |
| Labor market regulation | -0.05 | -0.06 | -0.02 | -0.009 |
| | (0.16) | (0.16) | (0.16) | (0.16) |
| Unemployment insurance | -0.12 | -0.07 | -0.30* | -0.25 |
| | (0.16) | (0.16) | (0.16) | (0.16) |
| Observations | 50 | 50 | 50 | 50 |
| Adjusted $R^2$ | 0.13 | 0.10 | 0.16 | 0.16 |

Note: *p<0.1; **p<0.05; ***p<0.01

## 5. Conclusion

This paper estimates panel FAVAR models to examine the state-level response of GDP, personal income, price levels, and employment to narrative U.S. federal tax changes. The findings uncover a substantial heterogeneity on the impact of personal and corporate income tax cuts on state-level macroeconomic variables. The magnitude and persistence of the GDP, personal income, and employment responses to either tax cut are estimated to be the largest in Colorado, Texas, Florida, California, Kansas, Arizona, and Nevada. In contrast, the federal personal income tax cut has a smaller impact on the GDP, personal income, and employment in states such as Oklahoma, Rode Island, New Mexico, Maine, New Hampshire, and Vermont. The FAVAR results highlight the significance of the New-Classical and the Keynesian models that link federal tax cuts with state-level real GDP and income growth. The findings of the FAVAR model are moderately consistent in directions and persistence with the conclusions of Marten and Ravn (2013) and Romer and Romer (2009).

Cross-state regression analysis suggests that the effect of tax cuts on real GDP, personal income, and employment are largest in states with a larger share of finance and manufacturing industries, higher nonfarm employment, and flexible supply of loans to small firms. States characterized by a higher amount of government debt, strict labor market regulations, a higher degree of economic policy uncertainty, and a higher tax burden appear to be negatively affected by federal tax changes. The cross-state regression analysis provides new insight into microeconomic channels of tax shock's transmission mechanism. The analysis broadly supports the economic significance of impulse response functions.

# 7. Appendix

## I. List of Bureau of Economic Analysis (BEA) regional macroeconomic variables

| Data/Series | Source |
|---|---|
| 1. Per Capita Personal Income in the Far West BEA Region<br>2. Personal Consumption Expenditures: Total for Far West BEA Region<br>3. Unemployment Rate in Far West BEA Region<br>4. Total Deposits in Commercial Banks in Far West BEA Region<br>5. Consumer Price Index for All Urban Consumers in Far West BEA Region<br>6. New Private Housing Units Authorized by Building Permit for the Far West BEA Region | 1. Federal Reserve Bank of St. Louis<br><br>2. BEA Regional Economic Accounts |
| 1. Per Capita Personal Income in the Great Lakes BEA Region<br>2. Personal Consumption Expenditures: Total for Great Lakes BEA Region<br>3. Unemployment Rate in Great Lakes BEA Region<br>4. Total Deposits in Commercial Banks in Great Lakes BEA Region<br>5. Consumer Price Index for All Urban Consumers in Great Lakes BEA Region<br>6. New Private Housing Units Authorized by Building Permit for the Great Lakes BEA Region | 1. Federal Reserve Bank of St. Louis<br><br>2. BEA Regional Economic Accounts |
| 1. Per Capita Personal Income in the Southwest BEA Region<br>2. Personal Consumption Expenditures: Total for Southwest BEA Region<br>3. Unemployment Rate in Southwest BEA Region<br>4. Total Deposits in Commercial Banks in Southwest BEA Region<br>5. Consumer Price Index for All Urban Consumers in Southwest BEA Region<br>6. New Private Housing Units Authorized by Building Permit for the Southwest BEA Region | 1. Federal Reserve Bank of St. Louis<br><br>2. BEA Regional Economic Accounts |
| 1. Per Capita Personal Income in the Southeast BEA Region<br>2. Personal Consumption Expenditures: Total for Southeast BEA Region<br>3. Unemployment Rate in Southeast BEA Region<br>4. Total Deposits in Commercial Banks in Southeast BEA Region<br>5. Consumer Price Index for All Urban Consumers in Southeast BEA Region<br>6. New Private Housing Units Authorized by Building Permit for the Southeast BEA Region | 1. Federal Reserve Bank of St. Louis<br><br>2. BEA Regional Economic Accounts |
| 1. Per Capita Personal Income in the Mideast BEA Region<br>2. Personal Consumption Expenditures: Total for Mideast BEA Region<br>3. Unemployment Rate in Mideast BEA Region<br>4. Total Deposits in Commercial Banks in Mideast BEA Region<br>5. Consumer Price Index for All Urban Consumers in Mideast BEA Region<br>6. New Private Housing Units Authorized by Building Permit for the Mideast BEA Region | 1. Federal Reserve Bank of St. Louis<br><br>2. BEA Regional Economic Accounts |
| 1. Per Capita Personal Income in the Rocky Mountain BEA Region<br>2. Personal Consumption Expenditures: Total for Rocky Mountain BEA Region<br>3. Unemployment Rate in Rocky Mountain BEA Region<br>4. Total Deposits in Commercial Banks in Rocky Mountain BEA Region<br>5. Consumer Price Index for All Urban Consumers in Rocky Mountain BEA Region | 1. Federal Reserve Bank of St. Louis<br><br>2. BEA Regional Economic Accounts |
| 6. New Private Housing Units Authorized by Building Permit for the Rocky Mountain BEA Region | |
| 1. Per Capita Personal Income in the Plains BEA Region<br>2. Personal Consumption Expenditures: Total for Plains BEA Region<br>3. Unemployment Rate in Plains BEA Region<br>4. Total Deposits in Commercial Banks in Plains BEA Region<br>5. Consumer Price Index for All Urban Consumers in Plains BEA Region<br>6. New Private Housing Units Authorized by Building Permit for the Plains BEA Region | 1. Federal Reserve Bank of St. Louis<br><br>2. BEA Regional Economic Accounts |
| 1. Per Capita Personal Income in the New England BEA Region<br>2. Personal Consumption Expenditures: Total for New England BEA Region<br>3. Unemployment Rate in New England BEA Region<br>4. Total Deposits in Commercial Banks in New England BEA Region<br>5. Consumer Price Index for All Urban Consumers in New England BEA Region<br>6. New Private Housing Units Authorized by Building Permit for the New England BEA Region | 1. Federal Reserve Bank of St. Louis<br><br>2. BEA Regional Economic Accounts |

## II. List of narrative tax changes (1977-2018)

| **Tax code/changes** | **Narrative PIT** | **Narrative CIT** |
|---|---|---|
| Tax Reduction and Simplification Act of 1977 | √ | √ |
| Revenue Act of 1978 | √ | √ |
| Economic Recovery Tax Act of 1981 | √ | √ |
| Deficit Reduction Act of 1984 | √ | √ |
| Tax Reform Act of 1986 | √ | √ |
| Omnibus Budget Reconciliation Act of 1987 | √ | √ |
| Omnibus Budget Reconciliation Act of 1990 | √ | √ |
| Omnibus Budget Reconciliation Act of 1993 | √ | **No** |
| Jobs and Growth Tax Relief Reconciliation Act of 2003 | √ | √ |
| The Tax Relief, Unemployment Insurance Reauthorization, and Job Creation Act of 2010 | √ | **No** |
| The Patient Protection and Affordable Care Act 2010 | √ | √ |
| The American Taxpayer Relief Act of 2012 | √ | √ |
| The Tax Cuts and Jobs Act of 2017 | √ | √ |

## III. List of state-level macroeconomic variables

| **Variable name** | **Source** | **FRED mnemonic** |
|---|---|---|
| Real GDP | FRED | RGSP |
| Disposable Personal Income | FRED | DPI |
| Total Nonfarm Payroll Employment | FRED | PAYEMS |
| Price Level | BLS | CPI |

## IV. List of U.S. States and abbreviations used in this study.

| State | Abbreviations | State | Abbreviations |
|---|---|---|---|
| Alabama | AL | Montana | MT |
| Alaska | AK | Nebraska | NE |
| Arizona | AZ | Nevada | NV |
| Arkansas | AR | New Hampshire | NH |
| California | CA | New Jersey | NJ |
| Colorado | CO | New Mexico | NM |
| Connecticut | CT | New York | NY |
| Delaware | DE | North Carolina | NC |
| Florida | FL | North Dakota | ND |
| Georgia | GA | Ohio | OH |
| Hawaii | HI | Oklahoma | OK |
| Idaho | ID | Oregon | OR |
| Illinois | IL | Pennsylvania | PA |
| Indiana | IN | Rhode Island | RI |
| Iowa | IA | South Carolina | SC |
| Kansas | KS | South Dakota | SD |
| Kentucky | KY | Tennessee | TN |
| Louisiana | LA | Texas | TX |
| Maine | ME | Utah | UT |
| Maryland | MD | Vermont | VT |
| Massachusetts | MA | Virginia | VA |
| Michigan | MI | Washington | WA |
| Minnesota | MN | West Virginia | WV |
| Mississippi | MS | Wisconsin | WI |
| Missouri | MO | Wyoming | WY |

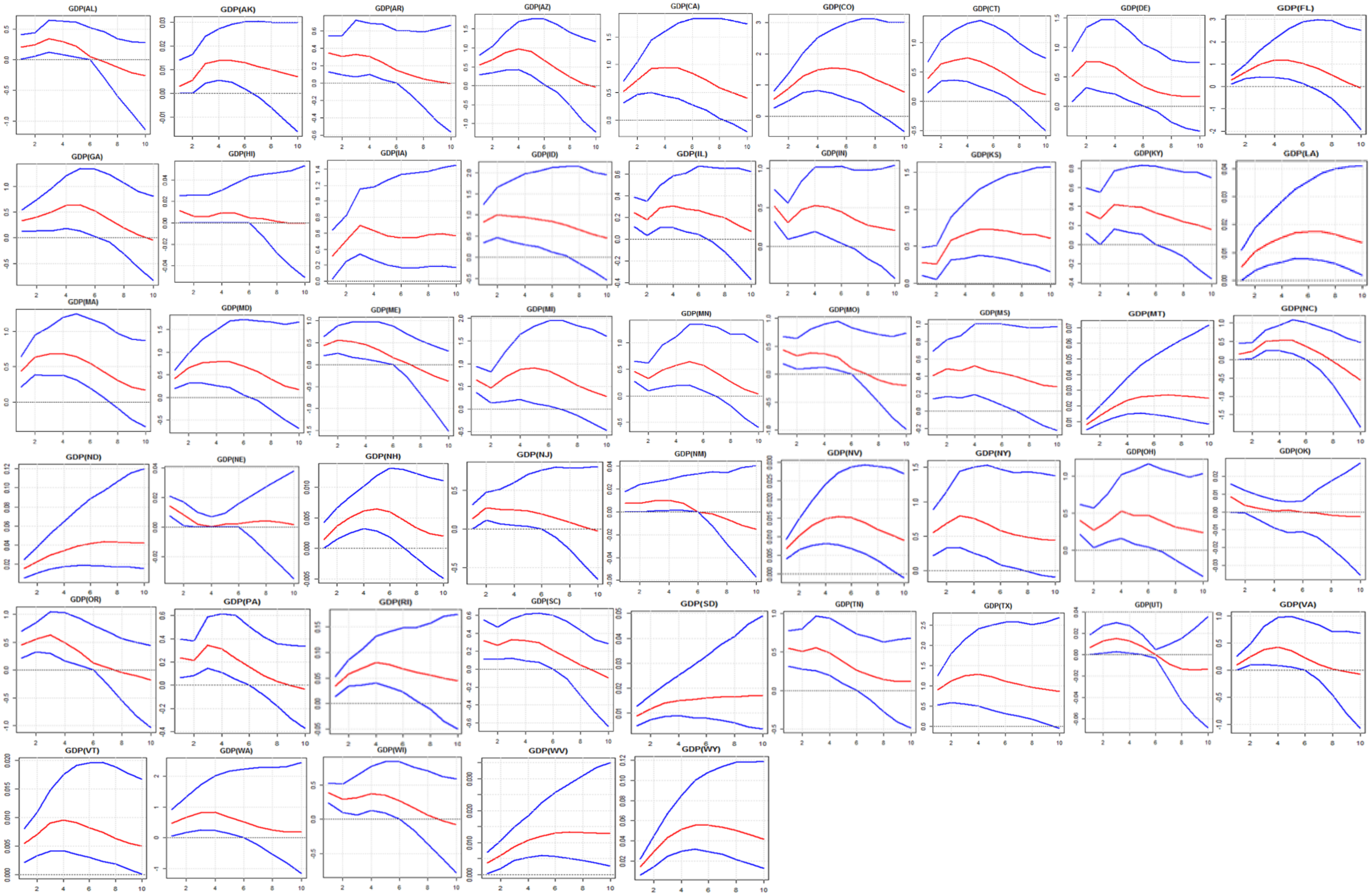

Figure 13: Impulse responses of state's GDP (PIT). IRFS to a 1 percent cut to the federal personal income tax rate.

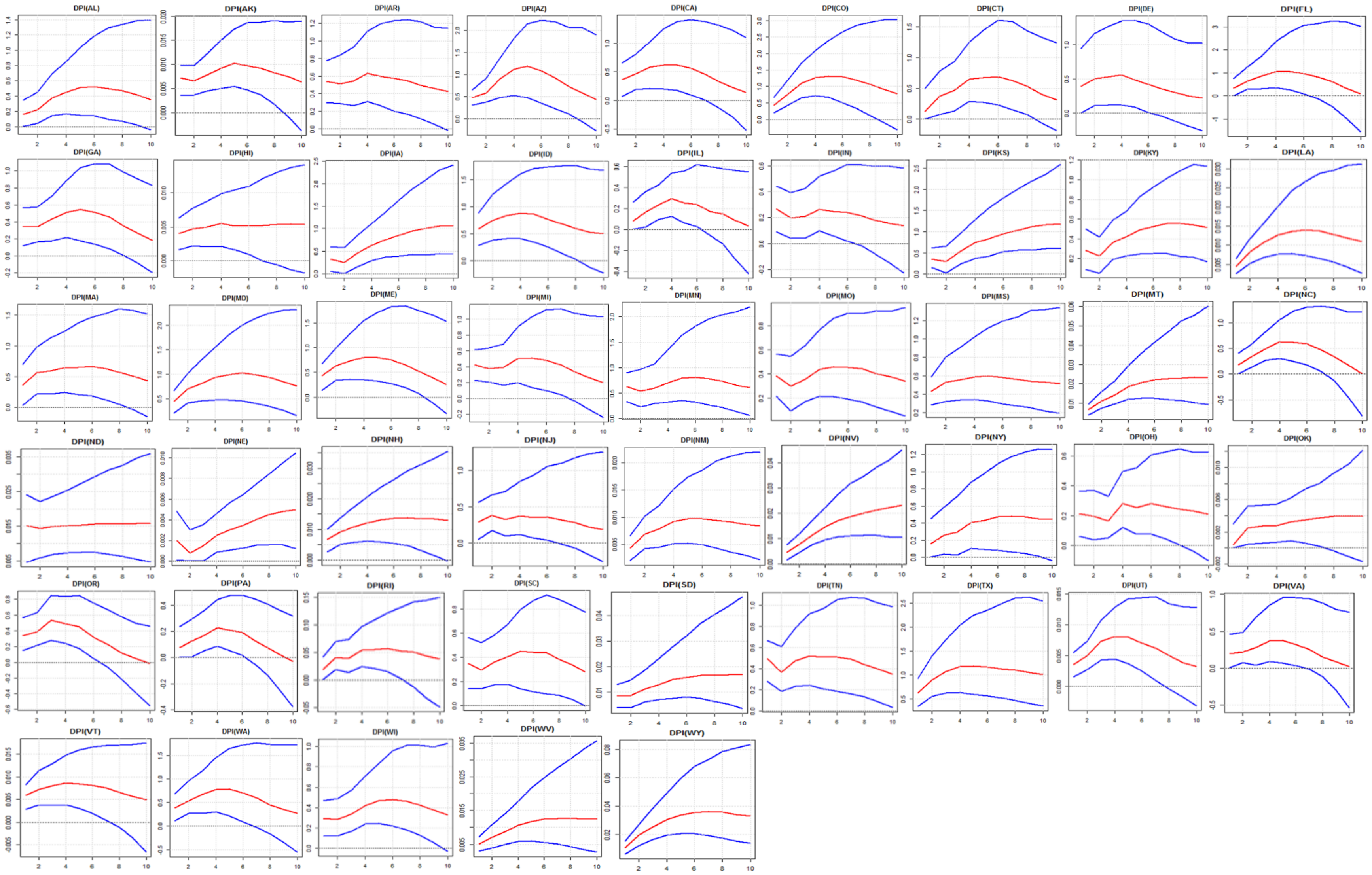

Figure 14: Impulse responses of state's DPI (PIT). IRFS to a 1 percent cut to the federal personal income tax rate.

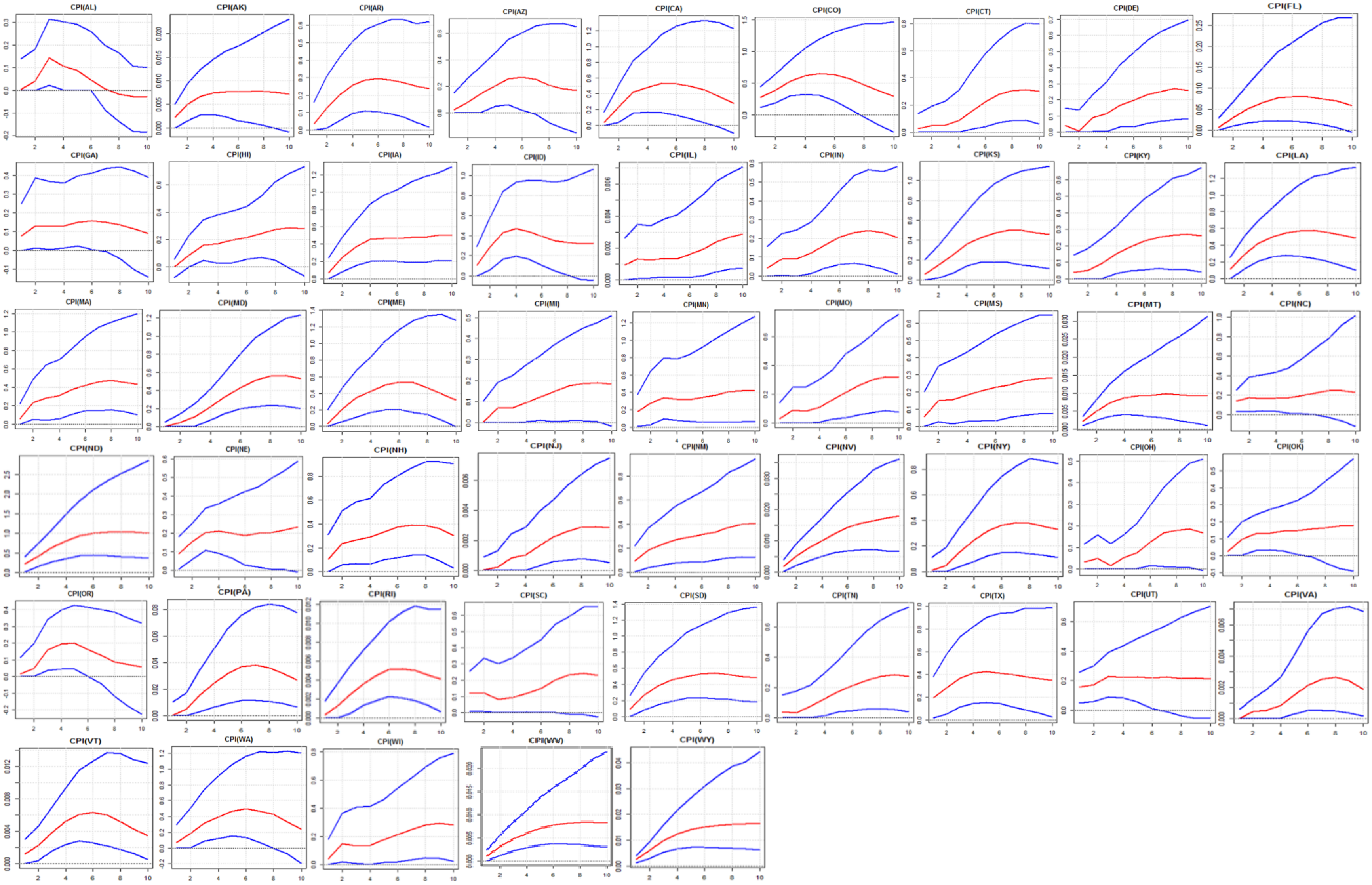

Figure 15: Impulse responses of state's CPI (PIT). IRFs to a 1 percent cut to the federal personal income tax rate.

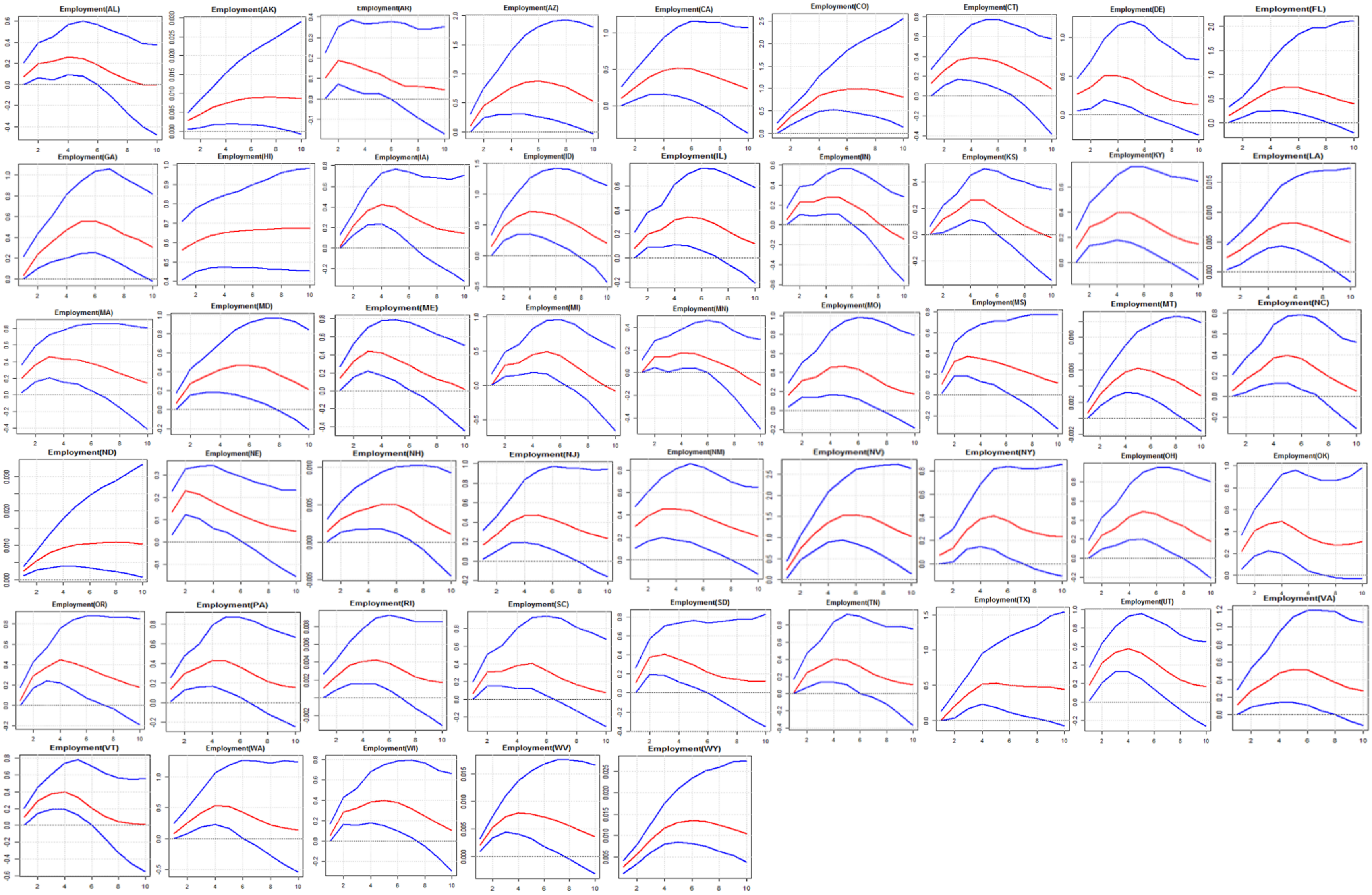

Figure 16: Impulse responses of state's employment (PIT). IRFs to a 1 percent cut to the federal personal income tax rate.

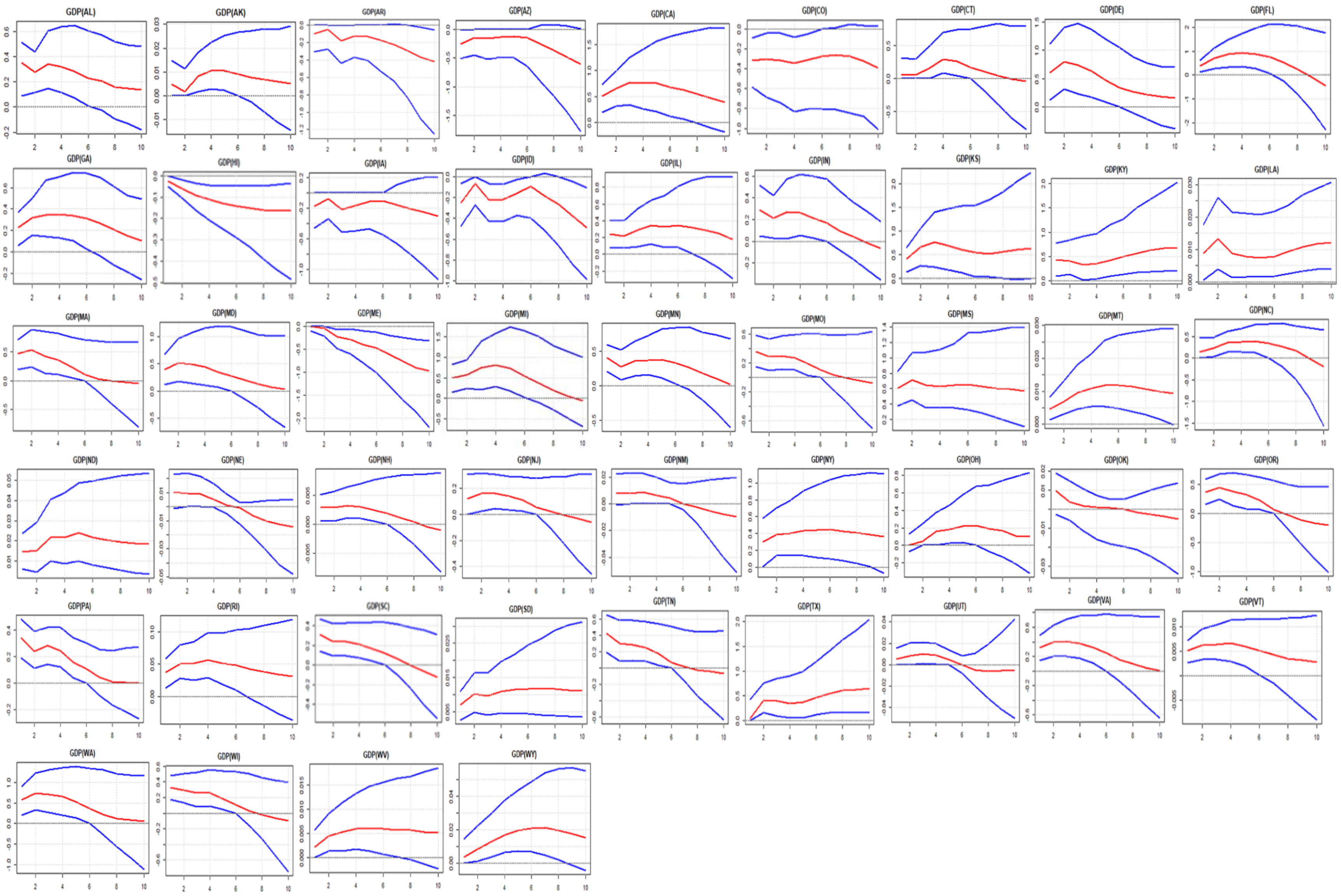

Figure 17: Impulse responses of state's GDP (CIT). IRFs to a 1 percent cut to the federal corporate income tax rate.

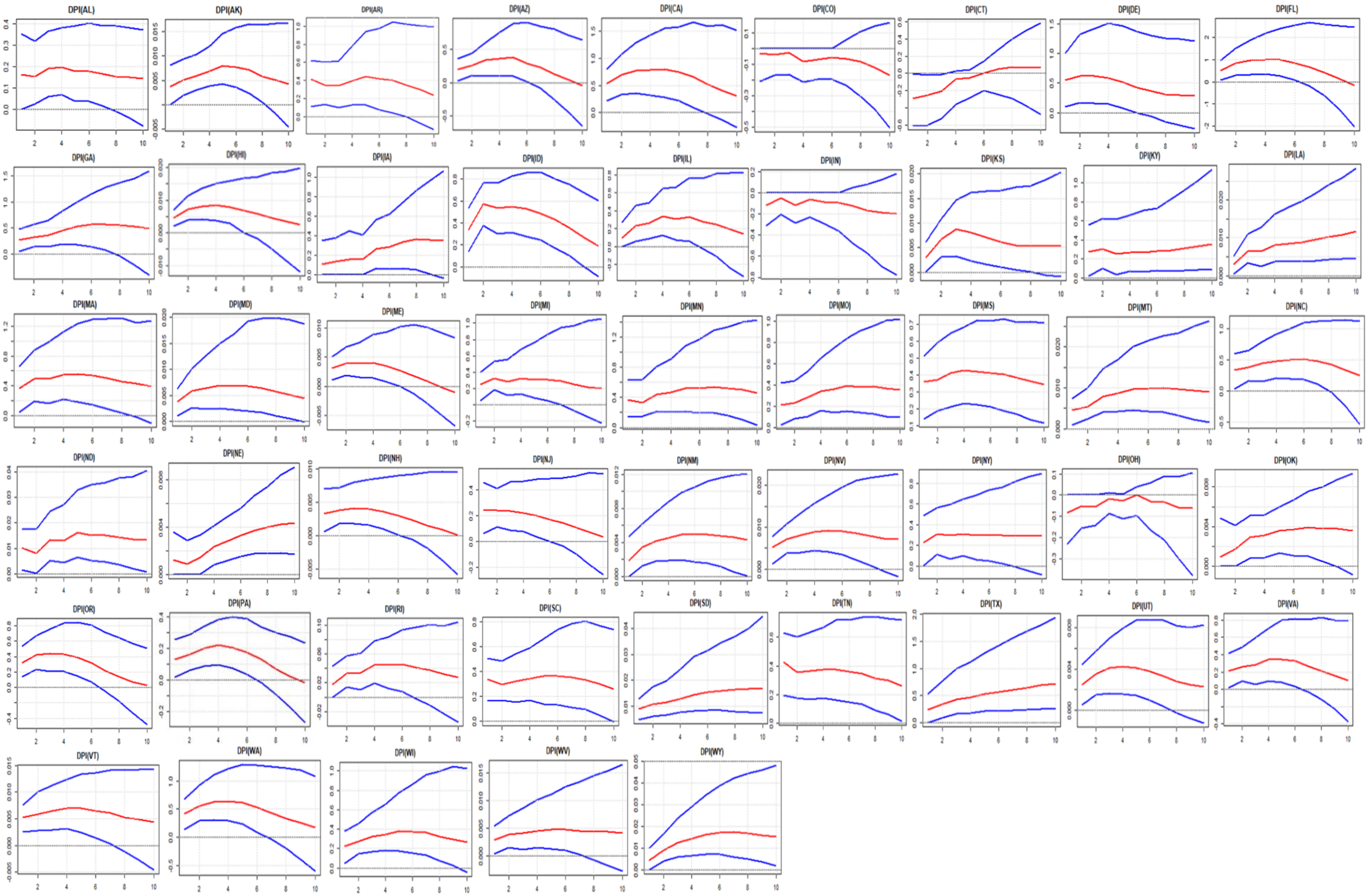

Figure 18: Impulse responses of state's DPI (CIT). IRFs to a 1 percent cut to the federal corporate income tax rate.

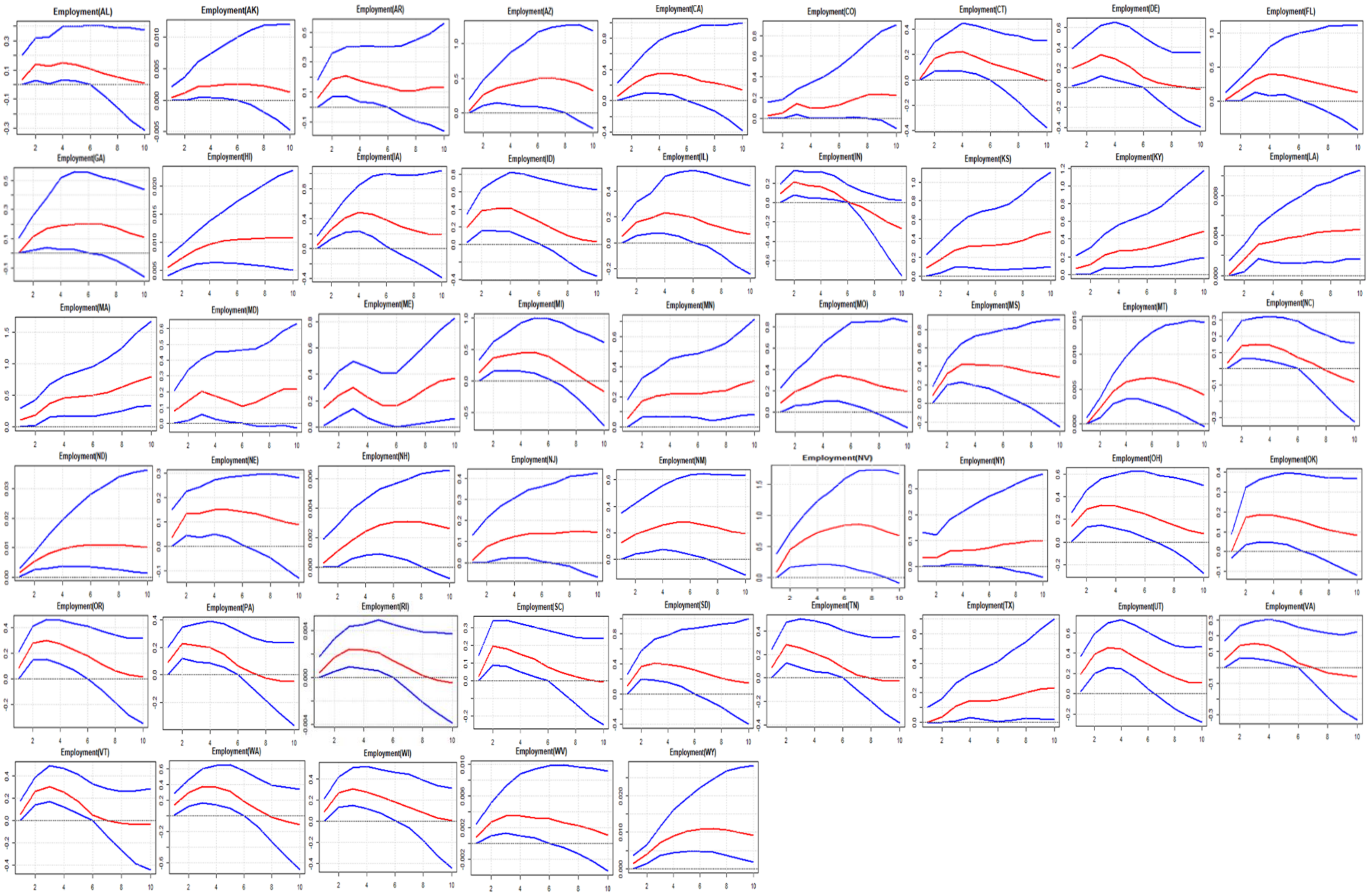

Figure 19: Impulse responses of state's employment (CIT). IRFs to a 1 percent cut to the federal corporate income tax rate.

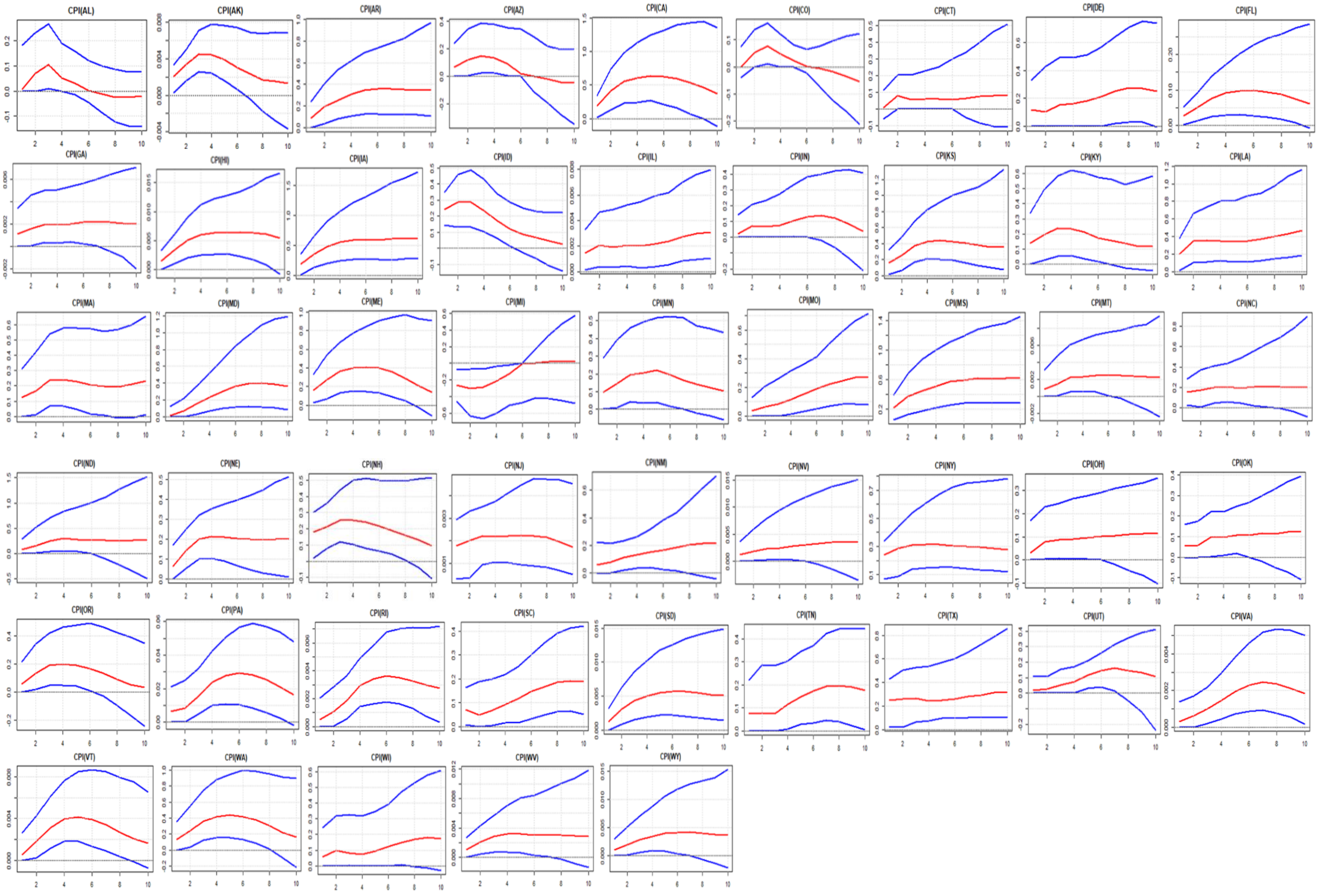

Figure 20: Impulse responses of state’s CPI (CIT). IRFs to a 1 percent cut to the federal corporate income tax rate.